\documentclass{article}
\usepackage{arxiv}

\usepackage[utf8]{inputenc} 
\usepackage[T1]{fontenc}    
\usepackage{hyperref}       
\usepackage{url}            
\usepackage{booktabs}       
\usepackage{amsfonts}       
\usepackage{nicefrac}       
\usepackage{microtype}      
\usepackage{graphicx}

\usepackage{afterpage}
\usepackage{amsmath}
\usepackage{amsfonts}
\usepackage{amssymb}
\usepackage{amsthm}
\usepackage{bm}
\usepackage{booktabs,multirow,multicol,makecell}
\usepackage[font=small,labelfont=small]{caption}
\usepackage{dsfont}
\usepackage{enumitem}
\usepackage{mathtools}
\usepackage{nicefrac}
\usepackage{parskip}
\usepackage{placeins}
\usepackage{newunicodechar,textgreek}
\usepackage{svg}
\usepackage{textgreek}
\usepackage{xcolor}

\usepackage{algorithm}
\usepackage{algpseudocode}

\usepackage{tikz}
\usetikzlibrary{tikzmark}
\usetikzlibrary{fit}

\usepackage[
    backend=biber,
    eprint=false,
    giveninits=true,
    maxcitenames=2,
    mincitenames=1,
    maxbibnames=10,
    minbibnames=1,
    sorting=none,
    style=phys,
    uniquename=init,
    doi=true,
    url=true
]{biblatex}

\DeclareFieldFormat{doi}{%
  DOI\addcolon\space
  \href{https://doi.org/#1}{\nolinkurl{#1}}%
}

\DeclareFieldFormat{url}{%
  URL\addcolon\space
  \url{#1}%
}

\AtEveryBibitem{%
  \clearlist{language}%
  \clearfield{month}%
  \clearfield{note}%
  \iffieldundef{doi}
    {}
    {\clearfield{url}}%
}

\usepackage{tabu}
\newcommand{\tblspace}{\vspace{1em}}

\newtheorem{theorem}{Theorem}

\makeatletter
\def\thm@space@setup{%
  \thm@preskip=\medskipamount
  \thm@postskip=\medskipamount
}
\makeatother

\newcommand{\idc}{\mathds{1}}

\usepackage{authblk}

\usepackage{mycommands}

\makeatletter
\g@addto@macro\textgreekfontmap{{ptm/m/n}{LGR/cmr/m/n}}
\makeatother
\makeatletter
\g@addto@macro\textgreekfontmap{{ptm/b/n}{LGR/cmr/m/n}}
\makeatother

\let\tablefontsize\small

\title{Quantifying network uncertainty in chemical reaction network inference}
\author[1,2]{Yong See Foo}
\author[1]{Adriana Zanca}
\author[1,2]{Jennifer A. Flegg}
\author[3]{Ivo Siekmann}
\affil[1]{School of Mathematics and Statistics, University of Melbourne, Australia}
\affil[2]{ARC Centre of Excellence for the Mathematical Analysis of Cellular Systems, University of Melbourne, Australia}
\affil[3]{School of Computer Science and Mathematics, Liverpool John Moores University, United Kingdom}
\date{}

\begin{document}

\maketitle

\begin{refsection}

\begin{abstract}

Dynamical systems in biology are complex, and one often does not have comprehensive knowledge about the interactions involved. Chemical reaction network (CRN) inference aims to identify, from time-course observations of species concentrations, the reactions occurring between species. Existing approaches such as sparse regularisation largely focus on identifying a single, most likely CRN, without addressing uncertainty about the network structure. However, it is important to quantify network uncertainty to have confidence in our inference and predictions. \review{In this work, we adapt sparse regularisation to quantify network uncertainty by mapping locally optimal solutions to CRN structures, which are subsequently weighted by approximate posterior probabilities. To validate our approach, we show in a simulation study that our method recovers CRNs that are structurally indistinguishable from the ground truth CRN. This involves extending theory on structural distinguishability to characterise the building blocks of structurally indistinguishable CRNs. We also find that nonconvex penalty functions give better coverage of plausible CRNs than the popular lasso regularisation. We also demonstrate that our method simultaneously recovers reactions proposed across multiple literature sources for a real-world reaction system.} Our emphasis on network-level probabilities enables a novel, hierarchical representation of reaction ambiguities, which highlights alternative reaction pathways supported by the data and can guide future experimental design.
\end{abstract}


\section{Introduction}

Reaction networks are frequently used to describe biological systems in fields ranging from cell biology to ecology \cite{resat_kinetic_2009,arceo_chemical_2015,warne_simulation_2019}. Chemical reaction networks (CRNs) are a useful framework for exploring biochemical systems in particular, where interactions between the chemical species are described as reactions. To construct a CRN that describes a dynamical system at hand, one typically posits a reaction network structure and subsequently estimates the associated parameters from experimental data. Parameter estimation is often accompanied by uncertainty and identifiability analyses, e.g., using profile likelihood methods \cite{raue_structural_2009} or Bayesian methods \cite{raue_joining_2013,hines_determination_2014}, which informs the reliability of the parameter inference results. However, these analyses mostly pertain to uncertainty that arises from parameter estimates; less attention has been paid to the uncertainty stemming from the specified network structure~\cite{galagali_exploiting_2019}. This does not reflect well the fact that one often does not have complete knowledge of the underlying species interactions. It is common wisdom that one cannot trust parameter estimates without uncertainty quantification; the same should be true for network inference. This motivates the problem of CRN inference which we tackle in this work: given a family of candidate CRNs, we seek to quantify the relative plausibility of the CRNs that reasonably explain time-course observations of the system.

A key challenge in CRN inference is that the combinatorial space of candidate CRNs is typically too large to enumerate exhaustively. Methods that tackle such CRN spaces can mostly be classified into sparse regularisation, stoichiometric optimisation, and model search. Sparse regularisation methods \cite{hoffmann_reactive_2019,gupta_parallel_2020,jiang_identification_2022} define a superstructure that consists of all candidate reactions with predefined stoichiometry, then apply regularisation during parameter inference to select a subset of reactions. Generally, this requires the user to tune a hyperparameter that balances model fit against model complexity. On the other hand, stoichiometric optimisation involves an objective function that assumes a fixed number of reactions of unknown stoichiometry, typically formulated with neural networks~\cite{ji_autonomous_2021,huang_data-driven_2022,li_bayesian_2023} or with integer programming~\cite{searsonReverseEngineering2012,willis_inference_2016,wilson_automated_2019}. Akin to hyperparameter tuning in sparse regularisation, the user is required to select the number of reactions. Finally, model search explicitly walks through the CRN space by iteratively evaluating candidate models, and is arguably the most complex in terms of implementation. The design of the model proposal must balance efficient exploitation of promising models and diverse exploration of the model space. Model search has been implemented under frameworks such as reversible-jump Markov chain Monte Carlo (RJMCMC) \cite{galagali_exploiting_2019}, bounded-depth tree search \cite{martinelli_reactmine_2023}, and genetic algorithms \cite{kreikemeyer_discovering_2024}.

These methods, which deal with combinatorially large CRN spaces, are mostly designed to infer a single CRN structure. For methods requiring hyperparameter tuning, even though different CRN structures are inferred over a range of hyperparameter values, considerable attention is given to selecting an ideal value to identify the ``best'' CRN structure. In data-limited settings, however, there may be insufficient information available to identify the correct CRN, a phenomenon known as practical indistinguishability~\cite{vajdaIdentifiabilityDistinguishability1994}. This issue of indistinguishability~\cite{godfreyProblemModel1989} is exacerbated by the fact that CRNs can be confoundable~\cite{craciun_identifiability_2008}---topologically distinct CRNs that can produce identical governing equations. We refer to this property as structural indistinguishability, which we elaborate in Section~\ref{sec:prelims_iden_dist}. In view of this uncertainty, predictions that are based on a single CRN structure can be overconfident and thus unreliable~\cite{burnhamModelSelection2004,chatfieldModelUncertainty1995}. CRN indistinguishability mirrors a well-studied issue at the parameter level: observed data are not always informative enough to discriminate between parameter values, a phenomenon known as parameter non-identifiability~\cite{erdiMathematicalModels1989,raue_structural_2009}. This work, however, is concerned with CRN distinguishability instead of parameter identifiability. Since CRNs cannot always be distinguished on the basis of observed data alone, rather than narrowing down a vast family of CRN structures to a single one, we instead account for network uncertainty by finding multiple CRN structures that each plausibly explain the observed data.

An important exception to the prevailing paradigm of inferring a single CRN structure is the Bayesian approach of RJMCMC \cite{galagali_exploiting_2019}, where the output of CRN inference is a posterior probability distribution over CRN structures. This reflects a Bayesian view of model selection \cite{kirk_model_2013}, where a prior belief about the model structure is updated in light of observed data, resulting in the posterior distribution. RJMCMC explores this posterior distribution by iteratively proposing model structures along with their corresponding parameters. When a model structure is proposed, the proposed parameter values need to provide a good fit to the data (relative to the current iteration) for the proposal to be accepted. If the acceptance rate is too low, RJMCMC fails to explore different model structures efficiently \cite{galagali_exploiting_2019,hastieModelChoice2012}. As noted previously, designing proposals that efficiently explore model space is a formidable challenge, especially for large, complex model spaces. 


Given the challenges regarding RJMCMC, we explore how effective sparse regularisation methods are for quantifying network uncertainty. In this work, network uncertainty is represented by a collection of CRNs, each weighted by a model probability. These CRNs are obtained from a collection of locally optimal solutions to a sparse regularisation problem. Such an approach does not guarantee that all plausible CRNs will be found via sparse regularisation; we find that the coverage of plausible CRNs depends on the regularisation induced by the penalty function chosen. Furthermore, we demonstrate that CRN coverage can be improved by a recombination strategy. We adopt a visual approach to illustrate reaction ambiguities by leveraging higher-order information in CRN uncertainty. Some of these identified ambiguities are corroborated by theory about structural distinguishability \cite{craciun_identifiability_2008}, but we also identify ambiguities that arise due to limited observed data, which cannot be predicted by theory alone. We demonstrate the applicability of our method to temporally sparse datasets of real chemical systems.


\section{Reaction network preliminaries}

\subsection{Mass-action kinetics}\label{sec:prelims_CRNs}

Consider a CRN of $S$ species, namely $X_1, \ldots, X_S$, whose concentrations we denote by 
$\mathbf{x} = (x_s)_{s=1}^S$. We define a complex $\cplx$ as a nonnegative integer combination of species, i.e. 
$\cplx = m_1X_1 + \cdots + m_SX_S$ where $m_1,\ldots,m_S\ge 0$ are assumed to be known. Let $\cplxset$ be the set of all possible complexes. A CRN is a simple directed graph with the vertex set $\cplxset$. Each reaction is represented by a directed edge that connects a reactant complex to a product complex. Suppose a library of candidate reactions, denoted as $\fullCRN$, is available to us from domain knowledge about the underlying system. With $\cplxset$ and $\fullCRN$ fixed, a CRN can be identified by a subset $\CRN \subseteq \fullCRN$ of reactions, where we denote the reactant complex and the product complex of reaction $r\in \CRN$ as
\begin{equation*}
    \cplx^-_r = m_{r1}^-X_1 + \cdots + m_{rS}^-X_S 
    \colswitch{\qquad \text{and} \qquad}{\quad \text{and} \quad}
    \cplx^+_r = m_{r1}^+X_1 + \cdots + m_{rS}^+X_S ,
\end{equation*}
respectively. We interchangeably let $\CRN$ refer to a set of reactions or the corresponding CRN, and let $\lvert\CRN\rvert$ denote the number of reactions in $\CRN$. The system of ordinary differential equations (ODEs) that govern the dynamics of $\mathbf{x}(t)$ induced by a CRN $\CRN$ are
\begin{equation}\label{eq:ode_system}
    \frac{dx_s}{dt} = \sum_{r\in\CRN} (m_{rs}^+ - m_{rs}^-)
    \rxrate_r(\mathbf{x}) \qquad \text{for } s=1,\ldots,S,
\end{equation}
where $\rxrate_r(\mathbf{x})$ is the rate of reaction $r\in\CRN$. Let $\mathbf{M}_\CRN$ denote the stoichiometric matrix of $\CRN$, which is a $\lvert \CRN \rvert \times S$ matrix whose $(r,s)$-th entry is ${m_{rs}^+ - m_{rs}^-}$, and let $\mathbf{m}_r$ denote the column corresponding to reaction~$r$, which is known as the reaction vector of reaction~$r$. We rewrite \eqref{eq:ode_system} as
\begin{equation}\label{eq:odemat}
    \frac{d\mathbf{x}}{dt} = \mathbf{M}_\CRN \Rxrate_\CRN(\mathbf{x}),
\end{equation}
where $\Rxrate_\CRN(\mathbf{x}) = (\rxrate_r(\mathbf{x}))_{r\in\CRN}$. Assuming deterministic mass-action kinetics, the rate of reaction $r$ is given by
\begin{equation}\label{eq:reaction_rate}
    \rxrate_r(\mathbf{x}) = k_r \prod_{s=1}^S x_{s}^{m_{rs}^-},
\end{equation}
where $k_r$ is the rate constant for reaction $r$.

\subsection{Identifiability and distinguishability}\label{sec:prelims_iden_dist}
A common task in CRN modelling is to estimate the rate constants, which we refer to as parameters, given observations of the concentrations~$\mathbf{x}$. However, there is no guarantee that these parameters can be estimated unambiguously, a phenomenon known as parameter non-identifiability \cite{erdiMathematicalModels1989,raue_structural_2009}. Literature on parameter identifiability often distinguishes between structural non-identifiability and practical non-identifiability: the former is concerned with non-identifiability
which can be explained by the model structure alone; the latter is concerned with non-identifiability due to insufficient information in noisy observations that are limited in quantity or quality. Assuming that all species are observed, we consider a deterministic mass-action CRN to be structurally non-identifiable if it has two parameterisations, i.e. two distinct sets of positive rate constants, that induce identical dynamics as defined in~\eqref{eq:odemat}--\eqref{eq:reaction_rate}. The structural identifiability of deterministic mass-action CRNs is characterised by Theorem~3.2 of \textcite{craciun_identifiability_2008}. On the other hand, the assessment of practical identifiability is more subjective; see e.g., the profile likelihood method \cite{raue_structural_2009}. Parameter identifiability presupposes that the network structure itself is known and fixed; assessing it is therefore beyond the scope of this work, which instead addresses the uncertainty of the network structure given data.

Adjacent to parameter identifiability is the concept of model distinguishability \cite{godfreyProblemModel1989,vajdaIdentifiabilityDistinguishability1994}, that is, the inability to distinguish two model structures on the basis of observed data alone. The possibility of CRN indistinguishability motivates the need to account for network uncertainty in CRN inference. As with parameter identifiability, model distinguishability can manifest as a structural or practical issue \cite{vajdaIdentifiabilityDistinguishability1994}. Assuming that all species are observed, we consider two deterministic mass-action CRNs to be structurally indistinguishable if there are parameterisations of each CRN that induce identical dynamics. \review{Structurally indistinguishable CRNs differ in their reaction sets, but do not necessarily differ in the complexes involved.} On the other hand, two CRNs are practically indistinguishable if the data do not significantly favour one CRN over the other due to insufficient information in noisy observations that are limited in quantity or quality.

Given two deterministic mass-action CRNs, one can test whether they are structurally indistinguishable via Theorem~4.4 of \textcite{craciun_identifiability_2008}, which we state here. Let $\textrm{Cone}(\cplx; \CRN)$ denote the open convex cone generated by the reaction vectors whose reactant complex is $\cplx$, namely
\begin{equation*}
    \textrm{Cone}(\cplx; \CRN) = \Biggl\{ \sum_{\substack{r\in \CRN \\ \cplx^-_r = \cplx}} u_r \mathbf{m}_r \colon u_r > 0 \text{ if } \cplx^-_r = \cplx \Biggr\}.
\end{equation*}
Note that if no reaction $r\in\CRN$ has $\cplx^-_r = \cplx$, then $\textrm{Cone}(\cplx; \CRN)$ consists of only the zero vector \cite{faulIdentifiabilitySDEs2026,szederkenyiCommentIdentifiability2009}.

\begin{theorem}[\textcite{craciun_identifiability_2008}]\label{thm:distinguish}
Under deterministic mass-action kinetics, two CRNs $\CRN$ and $\CRN'$ are structurally indistinguishable if and only if $\textrm{Cone}(\cplx; \CRN) \cap \textrm{Cone}(\cplx; \CRN')$ is nonempty for each complex~$\cplx$.
\end{theorem}
This result holds even if $\CRN$ corresponds to an empty set of reactions, which induces constant dynamics. For example, we consider $\{X_1\rightarrow\varnothing,X_1\rightarrow 2X_1\}$ to be structurally indistinguishable from the empty set of reactions, since assigning equal rate constants for $X_1\rightarrow\varnothing$ and $X_1\rightarrow 2X_1$ results in constant dynamics. Since structural distinguishability is a data-agnostic property, testing whether two CRNs are structurally distinguishable does not require measured observations. In Section~\ref{sec:elementary}, we describe how, given the library~$\fullCRN$ of candidate reactions, one can determine the building blocks from which every pair of structurally indistinguishable CRNs can be derived.

\section{Methods}

We now turn to the problem of inferring the CRN structure given time-course observations of its species concentrations. Since structurally or practically indistinguishable CRNs cannot be resolved on the basis of observations alone, our aim is not to identify a single best-fitting network, but rather quantify the uncertainty of a collection of CRNs that plausibly explain the data.

\subsection{Parameter inference}\label{sec:param_infer}
Suppose that we have a ground-truth CRN consisting of reactions $\CRN^\text{true}\subseteq\fullCRN$ with corresponding rate constants.
Our aim is to infer the CRN structure along with the rate constants from noisy observations of the concentrations $\mathbf{x}(t)$. We observe the data ${\data = (\mathbf{y}_n)_{n=1}^N}$ at time points $(t_n)_{n=1}^N$, where ${\mathbf{y}_n = \mathbf{x}(t_n) + \bm\xi_n}$ for ${n=1,\ldots,N}$, and $(\bm\xi_n)_{n=1}^N$ are independently distributed noise terms. For simplicity, we assume for each $n=1,\ldots,N$ that $\bm\xi_n$ follows an uncorrelated normal distribution with zero mean and variances $\bm{\sigma}^2 = (\sigma^2_s)_{s=1}^S$.

During inference, we assume that we know the library of candidate reactions $\fullCRN$ and the initial state of the dynamical system, $\mathbf{x}(0)$. Our approach is to estimate from data the rate constants $\mathbf{k}=(k_r)_{r\in\fullCRN}$ of the full CRN consisting of all candidate reactions $\fullCRN$, with the expectation that $k_r$ should be zero for reactions $r \in \fullCRN \setminus \CRN^\text{true}$. The noise variances $\bm{\sigma}^2$ are estimated simultaneously with $\mathbf{k}$. Let $\tilde{\mathbf{x}}(t;\mathbf{k},\mathbf{x}(0))\in\mathbb{R}^S$ denote the state at time $t$ of the full CRN with rate constants $\mathbf{k}$, which is obtained by solving the system of ODEs \eqref{eq:odemat}--\eqref{eq:reaction_rate} with a known initial state $\mathbf{x}(0)$, where $\CRN=\fullCRN$. The negative log-likelihood of the parameters $(\mathbf{k},\bm{\sigma}^2)$ is 
\begin{equation}\label{eq:neg_loglike}
\colswitch{
    -\log p(\data\vert\mathbf{k},\mathbf{x}(0),\bm{\sigma}^2) = \frac{1}{2} \sum_{s=1}^S \sum_{n=1}^N \left( \log(2\pi\sigma_s^2) +  \frac{(y_{ns} - \tilde{x}_s(t_n;\mathbf{k},\mathbf{x}(0)))^2}{\sigma_s^2} \right),
}{
\begin{split}
    {}&-\log p(\data\vert\mathbf{k},\mathbf{x}(0),\bm{\sigma}^2) \\
    ={}&\frac{1}{2} \sum_{s=1}^S \sum_{n=1}^N \left( \log(2\pi\sigma_s^2) +  \frac{(y_{ns} - \tilde{x}_s(t_n;\mathbf{k},\mathbf{x}(0)))^2}{\sigma_s^2} \right),
\end{split}
}
\end{equation}
where $y_{ns}$ and $\tilde{x}_s$ denote the $s$-th entry of $\mathbf{y}_n$ and $\tilde{\mathbf{x}}$, respectively. For conciseness, we suppress the dependence on $\mathbf{x}(0)$ hereafter.

Parameter estimation via minimising the negative log-likelihood \eqref{eq:neg_loglike} over $\bm\theta \coloneqq (\mathbf{k},\bm{\sigma}^2)$ is prone to overfitting \cite{hoffmann_reactive_2019}, as the number of reactions in $\CRN^\text{true}$ is typically much smaller than the number of all candidate reactions. We thus use a regularisation strategy to counteract this. Specifically, we minimise the loss function 
\begin{equation}\label{eq:loss_func}
    l(\bm\theta;\lambda) = -\log p(\data\vert\mathbf{k},\bm{\sigma}^2) + \sum_{r \in \fullCRN} \text{pen}(k_r; \lambda)
\end{equation}
over $\bm\theta$, where $\text{pen}(k;\lambda)$ is a penalty function that penalises larger values of the rate constant $k$, and $\lambda$ is a hyperparameter that controls the penalty strength. This approach can be interpreted as performing maximum \textit{a posteriori} (MAP) estimation, where the penalty function is understood to be the negative log prior density up to an additive constant. In this work, we investigate four penalty functions:
\begin{itemize}
    \item \emph{$L_1$ penalty:} $\text{pen}(k; \lambda) = \lambda k$. This is otherwise known as lasso regularisation, and is arguably the most common penalty function in sparse identification of nonlinear dynamics, e.g., \textcite{hoffmann_reactive_2019,pfister_learning_2019,egan_automatically_2024}.

    \item \emph{Log-scale $L_1$ penalty:} $\text{pen}(k; \lambda) = \lambda \lvert \log k - \log \varepsilon \rvert$ for some $0 < \varepsilon \ll 1$. Motivated by the fact that CRNs may feature reactions that act on different timescales, an $L_1$ penalty on a shifted log scale provides equal regularisation strength across different scales of the rate constant $k$. This effectively regularises $\log k$ towards a large negative value $\log \varepsilon$, as done in \textcite{gupta_parallel_2020}. For this, we use a value of $\varepsilon=10^{-10}$.

    \item \emph{Approximate $L_0$ penalty:} $\text{pen}(k; \lambda) = \lambda k^\rho$ for some $0 < \rho \ll 1$. This penalty function aims to penalise the log-likelihood according to the number of non-negligible parameters. Following \textcite{santosa_inverse_2011}, we choose $k^\rho$ as a continuous, differentiable approximation of the $L_0$ norm. In this work, we use a value of $\rho=0.1$.

    \item \emph{Horseshoe-like penalty:} $\text{pen}(k; \lambda) = -\log(\log (1+1/(\lambda k)^2))$. The horseshoe prior is a popular choice in sparse Bayesian estimation \cite{carvalho_horseshoe_2010}, and has been used in CRN inference \cite{jiang_identification_2022}. This penalty function is the negative log density of the horseshoe-like prior \cite{bhadra_horseshoe-like_2021}, which is a closed-form approximation to the original horseshoe prior. 
\end{itemize}

We obtain estimates of $\bm\theta$ by finding multiple local minima of the loss function $l(\bm\theta;\lambda)$. This is because we expect multiple CRN structures to be capable of plausibly explaining our observed data, which should correspond to local minima induced by our penalty function. We run multiple instances of the Broyden–-Fletcher–-Goldfarb–-Shanno (BFGS) algorithm \cite{nocedal_quasi-newton_2006} with $N_\text{hyp}$ different values of the penalty hyperparameter $\lambda$, each with $N_\text{start}$ runs from different starting points. BFGS is a gradient-based local optimisation algorithm that terminates upon reaching a local minimum. Varying the penalty hyperparameter allows us to find CRN structures with different model complexities, while varying the starting points allows us to access basins of attraction of different local minima. We retain all optimisation solutions across the $N_\text{hyp}N_\text{start}$ runs for each penalty function.

Note that real systems may exhibit reactions that occur on different timescales. Thus, instead of optimising over $\mathbf{k}$, we optimise over the transformed variables $\log (k_r/\kappa_r)$, where $r\in\fullCRN$ and $\kappa_r$ is a scaling factor determined \textit{a priori} of inference such that $k_r/\kappa_r$ is dimensionless. Moreover, instead of penalising each rate constant $k_r$ in \eqref{eq:loss_func}, we penalise the dimensionless parameter $k_r/\kappa_r$. As some of the penalty functions are undefined at 0, we impose a lower bound of ${k_r/\kappa_r \ge \varepsilon = 10^{-10}}$. We apply a log transformation following the advice of \textcite{villaverde_benchmarking_2019}, who found that optimisation on the log scale outperforms optimisation on the linear scale when estimating parameters for kinetic models. Details of our optimisation procedure are reported in Supplementary Materials~S1.

\subsection{Mapping parameter estimates to CRNs}\label{sec:map_ests}

In the existing literature on CRN inference, an estimate $\hat{\mathbf{k}}$ is mapped to a CRN by selecting reactions whose estimated rate constants exceed some threshold. This threshold is either chosen manually by visual inspection \cite{hoffmann_reactive_2019,kreikemeyer_discovering_2024,santosa_inverse_2011}, or chosen by searching through a set of thresholds such that the model fit with the selected parameters only is similar to the model fit with all parameters included \cite{jiang_identification_2022,ji_autonomous_2021}. However, for the sake of quantifying structural uncertainty, we seek to identify an ensemble of CRNs that can plausibly explain the data, based on the collection of $N_\text{hyp}N_\text{start}$ estimates found (with some penalty function) during parameter inference, which we denote as $\estset$. Note that a separate collection of estimates $\estset$ is obtained with each penalty function, implying that the CRN ensembles obtained with each penalty function are in general different. These CRN ensembles are constructed in two stages: a pruning stage and a recombination stage. Here, we outline the ideas behind the two stages; see Supplementary Materials~S2 for details. 

During the pruning stage, for each estimate $(\hat{\mathbf{k}}, \hat{\bm\sigma}^2) \equiv \est\in\estset$, we determine which reactions contribute significantly to the system dynamics induced by $\est$. The contribution of reaction $r$ is quantified by
\begin{equation}\label{eq:reaction_contribution}
    g(r;\hat{\mathbf{k}}) = \int_0^{t_N} \hat{k}_r \prod_{s=1}^S \tilde{x}_s(t;\hat{\mathbf{k}})^{m_{rs}^-} \, dt.
\end{equation}
Given that $\hat{\mathbf{k}}$ was obtained with sparse optimisation, we prune reactions with negligible values of $g(r;\hat{\mathbf{k}})$, thus mapping $\est$ to the CRN consisting of the remaining reactions. The pruning threshold is selected by ensuring that the system dynamics induced by the remaining reactions and the original estimate $\est$ are similar enough (Supplementary Materials~S2.1). Mapping the collection of estimates $\estset$ results in a base set of CRNs, which we denote as $\CRNset_\text{base}(\estset)$. However, since it is unlikely that all local minima of \eqref{eq:loss_func} are found during parameter inference, we expect $\CRNset_\text{base}(\estset)$ to be missing some CRNs that can plausibly explain the data. The recombination stage aims to recover such CRNs by combining CRNs from $\CRNset_\text{base}(\estset)$. Suppose that $\CRN^1,\CRN^2\in\CRNset_\text{base}(\estset)$ are two CRNs found in the pruning stage that are highly similar, in the sense that they share most of their reactions. The sets of reactions ${\CRN^{1\setminus 2}\coloneqq\CRN^1\setminus \CRN^2}$ and ${\CRN^{2\setminus 1}\coloneqq\CRN^2\setminus \CRN^1}$ may potentially contribute similarly to the system dynamics. The recombination stage involves finding appropriate pairs of CRNs $\CRN^1,\CRN^2\in\CRNset_\text{base}(\estset)$ such that reactions $\CRN^{2\setminus 1}$ can be substituted by $\CRN^{1\setminus 2}$, and applying these substitutions to the CRNs in $\CRNset_\text{base}(\estset)$. For each CRN $\CRN\in\CRNset_\text{base}(\estset)$ such that $\CRN^{2\setminus 1} \subseteq \CRN$, we propose the CRN ${(\CRN\setminus\CRN^{2\setminus 1})\cup\CRN^{1\setminus 2}}$, as it potentially results in dynamics similar to those induced by $\CRN$. Let $\CRNset(\estset)$ denote the ensemble of CRNs obtained across both stages. The number of CRNs in $\CRNset(\estset)$ is much smaller than the number of all possible subsets of $\fullCRN$, such that the posterior distribution over $\CRNset(\estset)$ is computationally tractable, which we describe in Section~\ref{sec:posterior}. 

\subsection{Elementary pairs of structurally indistinguishable reaction sets}\label{sec:elementary}
We can optionally apply CRN distinguishability theory to find CRNs that are structurally indistinguishable from CRNs in $\CRNset_\text{base}(\estset)$ as promising candidates to include in $\CRNset(\estset)$. To do so, we introduce the notion of elementary pairs of structurally indistinguishable reaction sets, which act as building blocks of structurally indistinguishable CRNs. We define a pair of structurally indistinguishable reaction sets $(\CRN,\CRN')$ to be elementary if (i)~$\CRN$ and $\CRN'$ are not both empty, and (ii)~for any subsets $\CRN^*\subseteq \CRN$ and ${\CRN'}^*\subseteq \CRN'$ that are structurally indistinguishable and not both empty, we must have $(\CRN^*,{\CRN'}^*)=(\CRN, \CRN')$. For example, the pair of structurally indistinguishable reaction sets $(\{X_1\rightarrow\varnothing,X_1\rightarrow 2X_1\}, \{\})$ is elementary, since neither $\{X_1\rightarrow\varnothing\}$ nor $\{X_1\rightarrow 2X_1\}$ induces constant dynamics. Elementary pairs of structurally indistinguishable reaction sets are the building blocks of structurally indistinguishable CRNs due to the following decomposition result:

\begin{theorem}\label{thm:elementary}
    Every pair of structurally indistinguishable reaction sets $(\CRN, \CRN')$ can be written as a componentwise union of elementary pairs of structurally indistinguishable reaction sets $(\CRN_1, \CRN'_1), \ldots, (\CRN_n, \CRN'_n)$, i.e.
    \begin{equation*}
        (\CRN, \CRN') = \biggl( \bigcup_{i=1}^n \CRN_i, \bigcup_{i=1}^n \CRN'_i \biggr).
    \end{equation*}
\end{theorem}
We present a proof of Theorem~\ref{thm:elementary} in Supplementary Materials~S4.1. In brief, the theorem follows from the result that every vector in a linear subspace can be written as a conformal sum of elementary vectors of the subspace~\cite{rockafellarElementaryVectors1967}, where \emph{conformal} refers to a sign-matching property; \emph{elementary} refers to a support-minimal property. We provide in Supplementary Materials~S4.2 an algorithm to enumerate all elementary pairs of structurally indistinguishable reaction sets given a library~$\fullCRN$ of candidate reactions, based on the double description algorithm~\cite{motzkinDoubleDescription1953}. Note that this procedure does not make use of parameter estimation. Elementary pairs can potentially be used in the recombination stage of Section~\ref{sec:map_ests} to find plausible CRNs that were not encountered during sparse optimisation. However, we do not pursue this approach in this work. Instead, we reserve the use of elementary pairs for evaluation purposes: since elementary pairs characterise structural indistinguishability without relying on parameter estimation, they serve as an independent source of ground truth for testing the ability of sparse optimisation to recover structurally indistinguishable CRNs.

\subsection{Posterior distribution over CRNs}\label{sec:posterior}

To apply Bayesian model selection, we need to define a prior distribution over CRNs to describe our prior belief about the CRN. For each CRN $\CRN\subseteq\fullCRN$, let $p(\CRN)$ be the probability that $\CRN$ is the data-generating CRN, without knowledge of any observed data. The posterior distribution over CRNs represents an updated belief about the CRNs upon observing some dataset~$\data$. The probability that CRN $\CRN$ generates the dataset $\data$ is called the model evidence of CRN $\CRN$, denoted as $p(\data \vert \CRN)$. It follows from Bayes' theorem that the posterior probability of CRN $\CRN$ is given by
\begin{equation}\label{eq:bayes_crn}
    p(\CRN\vert \data) = \frac{p(\CRN)p(\data\vert \CRN)}{\sum_{\CRN' \subseteq \fullCRN} p(\CRN')p(\data\vert \CRN')}.
\end{equation}
Note that the model evidence $p(\data \vert \CRN)$, also known as marginal likelihood, does not feature rate constants, as they are integrated out. However, in the context of \eqref{eq:bayes_crn}, $p(\data \vert \CRN)$ should be interpreted as a \emph{likelihood} as we are performing inference at the level of CRN structures, not rate constants.

To specify a CRN prior $p(\CRN)$, we assume that each reaction is present in the network independently with a common probability $q$:
\begin{equation}\label{eq:iid_reactions}
    p(\CRN \vert q) = q^{\lvert\CRN\rvert}(1-q)^{\lvert\fullCRN\rvert-\lvert\CRN\rvert}.
\end{equation}
Fixing $q$ at some value would cause the false positive rate of spurious reactions to increase as the library size $\lvert\fullCRN\rvert$ grows~\cite{scott_bayes_2010}. Instead, we place a prior on $q$ to account for multiplicity correction~\cite{scott_bayes_2010}. A natural, uninformative prior on $q$ is the uniform distribution over $[0,1]$~\cite{geisserPriorDistributions1984}, yielding the following prior on CRNs:
\begin{equation}\label{eq:uniform_prior}
    p(\CRN) = \int_0^1  p(\CRN \vert q) \, dq  = \frac{1}{\lvert \fullCRN \rvert+1} \binom{\lvert \fullCRN \rvert}{\lvert \CRN \rvert}^{-1}.
\end{equation}
The prior~\eqref{eq:uniform_prior} implies that the number of reactions is \emph{a priori} uniformly distributed over $\{0,1,\ldots,\lvert \fullCRN \rvert\}$, reflecting a lack of prior preference for any particular network size. The uniform prior on $q$ is recognised as a default choice in variable selection literature~\cite{cuiEmpiricalBayes2008,scott_bayes_2010}, and has been used in CRN inference by \textcite{oates_causal_2014}. In general, one can replace the uniform prior on $q$ with a beta distribution \cite{cuiEmpiricalBayes2008} whose parameters can be informed by prior belief about the number of reactions in the network.

Next, we describe how we compute the model evidence of each CRN from the ensemble $\CRNset(\estset)$. Exact calculation of the model evidence involves an intractable integration over high-dimensional parameter space, so it is common to calculate the model evidence approximately; see \textcite{llorente_marginal_2023} for a review on model evidence computation. Here, we use the Bayesian information criterion (BIC) \cite{schwarz_estimating_1978}, given by
\begin{equation}\label{eq:bic}
\begin{split}
\colswitch
{
    \textsc{bic}(\CRN) &= -2 \hat{L} + \lvert \CRN \rvert \log \lvert \data \rvert,  \\
     \text{where } \hat{L} &= \max_{\bm\theta} \log p(\data\vert\mathbf{k},\bm\sigma^2) \quad \text{subject to } k_r = 0 \:\forall\, r \in \fullCRN\setminus\CRN 
}
{
    \textsc{bic}(\CRN) ={}&-2 \hat{L} + \lvert \CRN \rvert \log \lvert \data \rvert,  \\
     \text{where } \hat{L} ={}&\max_{\bm\theta} \log p(\data\vert\mathbf{k},\bm\sigma^2) \\
     {}&\text{subject to } k_r = 0 \:\forall\, r \in \fullCRN\setminus\CRN 
}
\end{split}
\end{equation}
and $\lvert\data\rvert$ denotes the number of observations. Although the multiplicative prefactor of $\log\lvert\data\rvert$ is usually defined as the number of model parameters, here we ignore the noise variances and count only the number of reactions, as the resulting discrepancy is an additive constant with respect to $\CRN$. For large $\lvert\data\rvert$, the BIC is approximately equal to $-2 \log p(\data\vert \CRN)$ \cite{schwarz_estimating_1978}. Note that the optimisation in \eqref{eq:bic} differs from minimisation of \eqref{eq:loss_func} as the penalty function from \eqref{eq:loss_func} is substituted with the constraint $k_r = 0 \,\forall\, r \in \fullCRN\setminus\CRN$ in \eqref{eq:bic}. Nevertheless, the estimates $\estset$ are used as starting values for the optimisation problems \eqref{eq:bic}; see Supplementary Materials~S2.3. We refer to the maximiser in \eqref{eq:bic} as the maximum likelihood estimate (MLE) under CRN $\CRN$.

An exact computation of the denominator of the posterior distribution~\eqref{eq:bayes_crn} requires summing over all possible CRNs $\CRN\subseteq\fullCRN$, which is computationally infeasible. Instead, we only sum over the CRNs in the ensemble $\CRNset(\estset)$, obtained from the collection of estimates $\estset$. Substituting the BIC approximation $\textsc{bic}(\CRN) \approx -2 \log p(\data\vert \CRN)$ gives us
\begin{equation}\label{eq:bayes_crn_sel}
    p(\CRN \vert \data) \approx \frac{p(\CRN) \exp (-\textsc{bic}(\CRN) /2 )}{\sum_{\CRN' \in \CRNset(\estset)} p(\CRN') \exp (-\textsc{bic}(\CRN') /2 )}
\end{equation}
for each $\CRN \in \CRNset(\estset)$. Since we obtain separate CRN ensembles $\CRNset(\estset)$ with each penalty function, the posterior distributions obtained with each penalty function are generally different. Note that the support of our approximate posterior~\eqref{eq:bayes_crn_sel} is a truncation of the support of the exact posterior~\eqref{eq:bayes_crn}. This truncation is appropriate only if the posterior mass is concentrated within $\CRNset(\estset)$, thus requiring the estimates $\estset$ to provide  parsimonious model fits that map to a variety of CRNs. In other words, the validity of \eqref{eq:bayes_crn_sel} depends on the quality and diversity of the estimates $\estset$, which in turn depends on the penalty function used. The comparison of penalty functions will be a key theme in our results. From here, references to the posterior refer to \eqref{eq:bayes_crn_sel} instead of \eqref{eq:bayes_crn}, unless stated otherwise. A summary of our proposed method is provided in Supplementary Materials, Algorithm S1. 

In Bayesian statistics, uncertainty is often reported as credible regions. Although credible regions are not unique, one can define the $100(1-\alpha)\%$ highest posterior density (HPD) region as the smallest region that contains at least $1-\alpha$ of the posterior probability mass. For CRN inference, we define the $100(1-\alpha)\%$ HPD set of CRNs to be the smallest subset of $\CRNset(\estset)$ such that the posterior probability of this subset of CRNs is at least $1-\alpha$. This subset is determined with a greedy approach, that is, we incrementally include the CRNs with the highest posterior probabilities until the sum of their posterior probabilities reaches $1-\alpha$. 

\section{Results}

\subsection{Simulation study: synthetic CRN}\label{sec:toy}

\begin{figure}
\includegraphics[width = 0.99\textwidth]{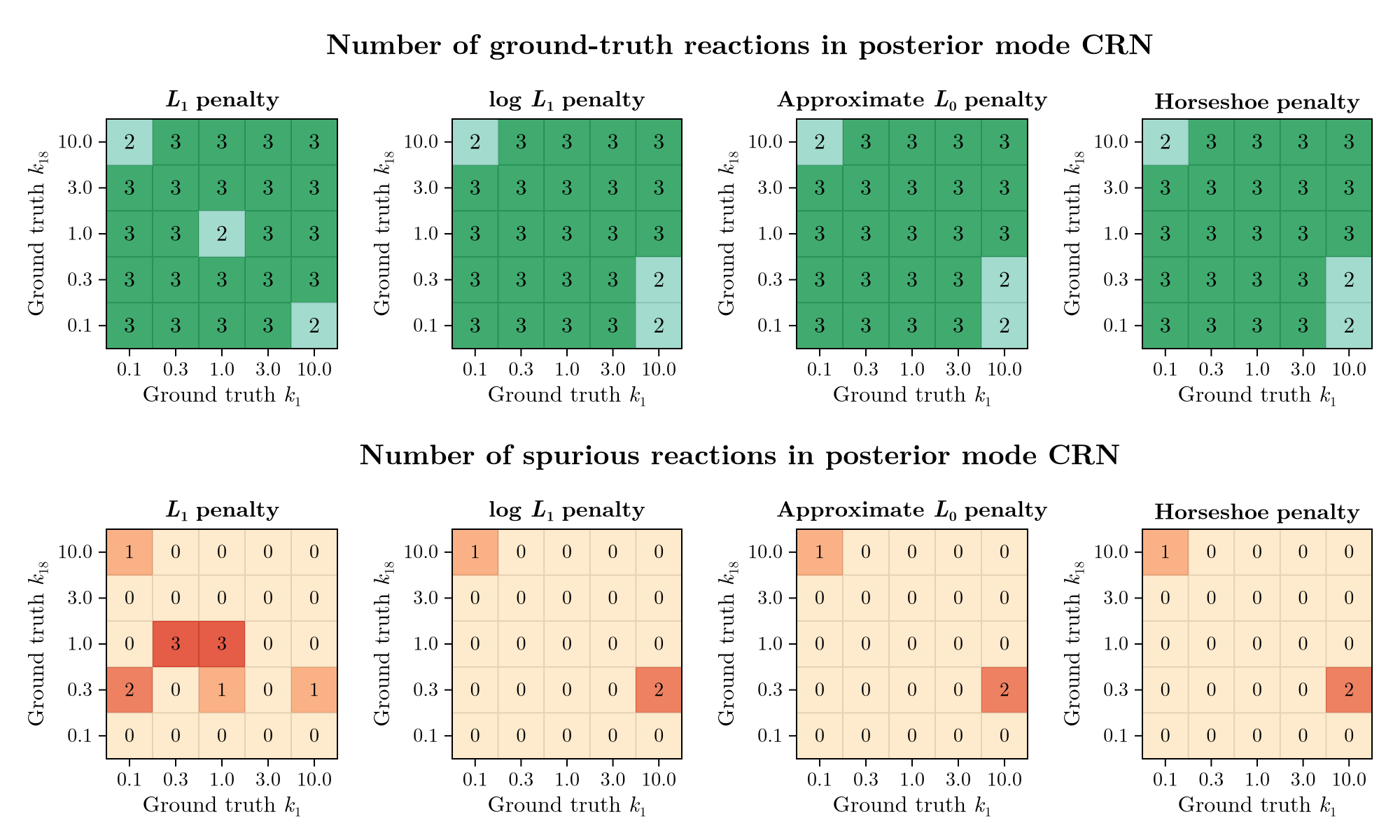}
\caption{Number of true positives (top) and false positives (bottom) in the set of reactions corresponding to the posterior mode CRN obtained with each penalty function for the simulation study.}
\label{fig:toy_mode_eval}
\end{figure}

\begin{figure}
\includegraphics[width = 0.99\textwidth]{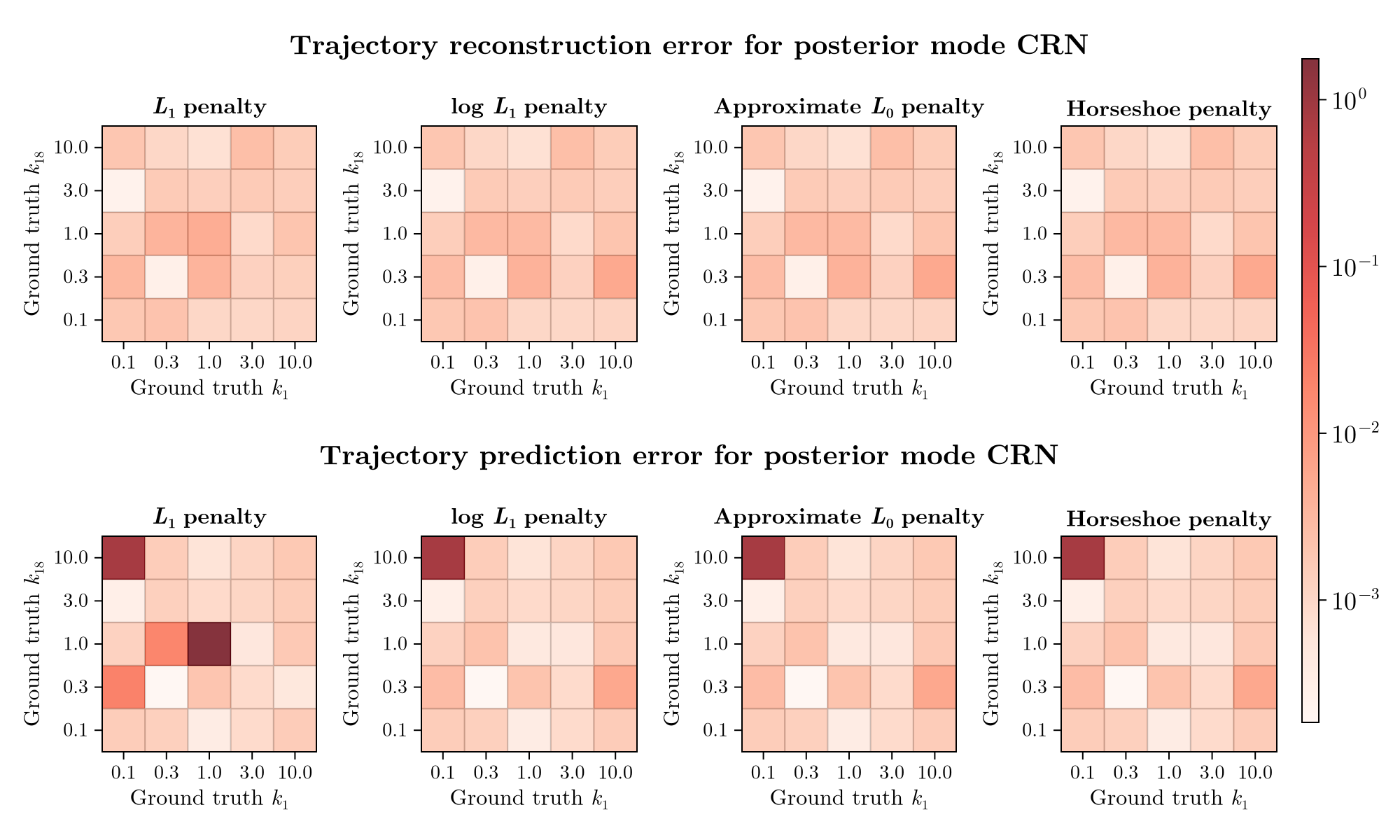}
\caption{Error of trajectories simulated using the MLE of the posterior mode CRN obtained with each penalty function, either from the original initial state $\mathbf{x}(0)=(0,0,1)$ (top) or a novel initial state $\mathbf{x}(0)=(1,0,0)$ (bottom). Error is computed as the time-averaged absolute errors between the ground-truth trajectories and the trajectories simulated using the MLE of the posterior mode CRN, summed over species; see Supplementary Materials~S5.1.}
\label{fig:toy_mode_traj}
\end{figure}

To illustrate our method, we consider data simulated from a synthetic CRN consisting of $S=3$ species and reactions
\begin{equation}\label{eq:toy_crn}
    X_1 \xrightarrow{k_1} X_2, \quad X_3 \xrightarrow{k_{13}} X_1+X_2, \quad X_1+X_2 \xrightarrow{k_{18}} X_3,
\end{equation}
where the rate constant indices follow the numbering of the candidate reactions (Supplementary Materials~S3.1). We simulate 25 datasets, each with different ground-truth rate constants (Figure~S1). Specifically, $k_1$ and $k_{18}$ take values in $\{0.1,0.3,1,3,10\}$, while $k_{13}=1$ is fixed. The initial state is set at ${\mathbf{x}(0) = (0,0,1)}$. For this simulation study, we are interested in the data regime where observations are densely sampled with minimal noise levels. Upon simulating the ODE trajectories, we sample observations at $N=101$ time points ${(0, 0.1, \ldots, 10)}$, where the added noise is normally-distributed with species-specific standard deviation equal to $1\%$ of the species' trajectory range.

We construct the library of candidate reactions $\fullCRN$ assuming that the possible complexes are \splitatcommas{$\cplxset = \{X_1,X_2,X_3,X_1+X_2,X_1+X_3,X_2+ X_3\}$}, and that reactions between all distinct pairs of complexes are possible. This results in a library of $6\times 5 = 30$ candidate reactions. We apply our CRN inference method under each of the four penalty functions (hereafter referred to as $L_1$, log $L_1$, approximate $L_0$, horseshoe), performing optimisation with ${N_\text{hyp}=10}$ hyperparameter values. For each dataset, penalty function, and hyperparameter value, we perform optimisation from ${N_\text{start}=64}$ starting points. See Supplementary Materials~S1 for choices of hyperparameter values and starting points. For each penalty function and dataset, we compute a separate posterior distribution (Section~\ref{sec:posterior}) based on the CRN ensemble $\CRNset(\estset)$ (Section~\ref{sec:map_ests}) obtained with that penalty function.


\subsubsection{Consequences of ignoring network uncertainty}\label{sec:consequences}

We first inspect the CRN with the highest posterior probability \eqref{eq:bayes_crn_sel}, which we call the posterior mode CRN, denoted as $\CRN^\text{mode}$. For each penalty function and dataset, we show in Figure~\ref{fig:toy_mode_eval} the number of reactions correctly and spuriously identified by $\CRN^\text{mode}$, i.e. true and false positives, respectively. In some cases, the convex $L_1$ penalty produces more spurious reactions than the other three penalty functions (which are nonconvex), despite a wide range of penalty hyperparameter values used (Supplementary Materials~S1.4). In cases where $\CRN^\text{mode}$ coincides with the ground-truth CRN $\CRN^\text{true}$ \eqref{eq:toy_crn}, we find that the ground-truth rate constants are recovered well by the MLE under $\CRN^\text{mode}$ (Figure~S2). However, we also seek to understand why for some datasets $\CRN^\text{mode}\neq\CRN^\text{true}$. For instance, consider the dataset where $k_1=0.1$ and $k_{18}=10$, featuring a slow ${X_1 \xrightarrow{k_1} X_2}$ reaction and a fast ${X_1+X_2 \xrightarrow{k_{18}} X_3}$ reaction. We find that for all penalty functions, $\CRN^\text{mode}$ differs from $\CRN^\text{true}$ by including the alternative reaction ${X_1+X_3 \xrightarrow{k_{30}} X_2+X_3}$ in place of the ground-truth reaction ${X_1 \xrightarrow{k_1} X_2}$. The dynamics of $X_3$ is biphasic (top left panel of Figure~S1): initially there is a brief rapid decrease in concentration, followed by a slow decline. The fact that $\CRN^\text{mode}$ involves the alternative reaction is likely due to an overfit to the data from the short transient phase, made possible because there is insufficient information from the remaining data, which stays relatively constant, to discriminate whether $X_3$ is a necessary catalyst for the conversion of $X_1$ to $X_2$. This is an example of practical indistinguishability as the reactant complexes of the two reactions, namely $X_1$ and $X_1+X_3$, are different.

Regardless of whether $\CRN^\text{mode}=\CRN^\text{true}$, simulating the posterior mode CRN using its corresponding MLE accurately reconstructs the original trajectories (Figure~\ref{fig:toy_mode_traj}, top row). However, if we use the MLE of the posterior mode CRN to predict trajectories starting from a novel initial state ${\mathbf{x}(0)=(1,0,0)}$, the resulting trajectories do not always match closely to the trajectories simulated with the ground-truth CRN (Figure~\ref{fig:toy_mode_traj}, bottom row). The posterior mode CRN obtained with the $L_1$ penalty results in poorer predictive power when it has spurious reactions, signifying an overfit to data. For the nonconvex penalty functions, the prediction errors are similar to the corresponding reconstruction errors, with the exception of the \splitatcommas{$k_1=0.1, k_{18}=10$} dataset. As noted before, $\CRN^\text{mode}$ for this dataset differs from $\CRN^\text{true}$ by including the alternative reaction ${X_1+X_3 \xrightarrow{k_{30}} X_2+X_3}$ in place of the ground-truth reaction ${X_1 \xrightarrow{k_{1}} X_2}$. Hence, under $\CRN^\text{mode}$, no conversion of $X_1$ to $X_2$ is possible as $X_3$ is absent in the novel initial state. Thus, regardless of the value of the rate constants, $\CRN^\text{mode}$ incorrectly predicts that no dynamics will occur, resulting in a large prediction error, irrespective of the parameter values.

\begin{figure}
\includesvg[width=\linewidth, inkscapearea=page, inkscapelatex=false]{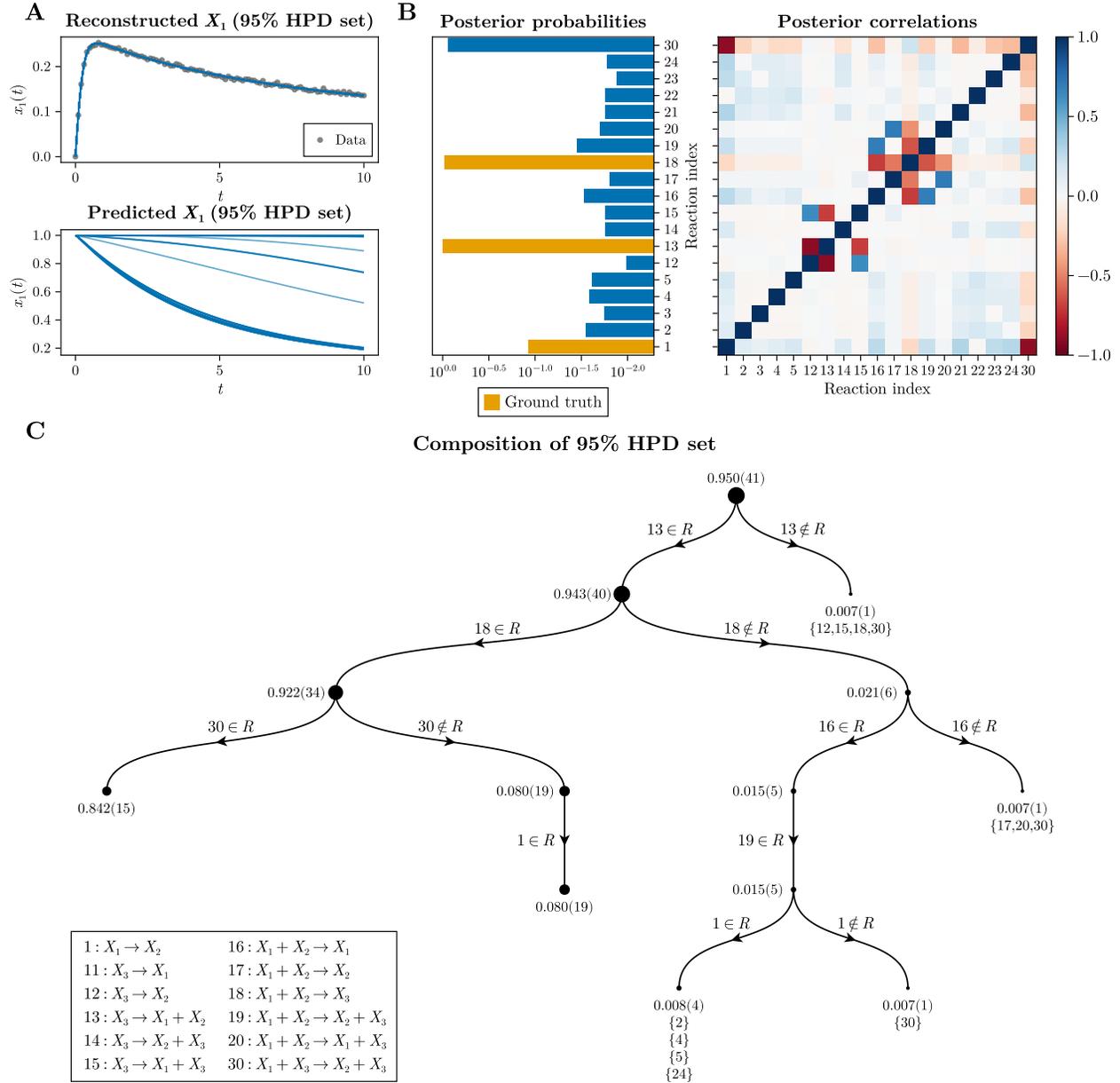}
\caption{Posterior summary for the $k_1=0.1,\,k_{18}=10$ dataset of the simulation study using the $\log L_1$ penalty. (A) Trajectories simulated with the MLEs of the CRNs in the 95\% HPD set either from the original initial state $\mathbf{x}(0)=(0,0,1)$ (top) or the novel initial state $\mathbf{x}(0)=(1,0,0)$ (bottom). (B) Posterior probabilities of reactions which are greater than $0.01$ (left), and their posterior correlations (right). (C) Hierarchical representation of the 95\% HPD set. The root node represents the 95\% HPD set, which is sequentially split according to reaction inclusion as indicated by the edges. Each node represents a subset of the 95\% HPD set, labelled with its posterior probability and number of CRNs in brackets. Node marker size scales with the number of CRNs. See main text for how splits are determined and details about the leaf nodes. The inset lists the reactions explicitly discussed in Section~\ref{sec:toy}.}
\label{fig:toy_uncertainty}
\end{figure}

\afterpage{\FloatBarrier}

\subsubsection{Accounting for uncertainty reveals alternative networks}

If we place all our confidence in the posterior mode CRN, we risk producing unreliable predictions without adequate uncertainty quantification. Instead, we account for network uncertainty by inspecting the rest of the posterior distribution over $\CRNset(\estset)$. We demonstrate this in Figure~\ref{fig:toy_uncertainty} using the posterior distribution obtained with the $\log L_1$ penalty for the \splitatcommas{$k_1=0.1,k_{18}=10$} dataset; other nonconvex penalty functions lead to similar results (not shown). We consider the 95\% HPD set to be a subset of CRNs which explain the observed data well. In this case, the 95\% HPD set consists of 41 CRNs, and the trajectories reconstructed using their corresponding MLEs all provide a good fit to the data (Figure~\ref{fig:toy_uncertainty}A, top panel). However, there is significant variation across the trajectories predicted from the novel initial state ${\mathbf{x}(0)=(1,0,0)}$ using the MLEs of the CRNs in the 95\% HPD set (Figure~\ref{fig:toy_uncertainty}A, bottom panel). Note that we have only shown the $X_1$ trajectories for visual clarity; similar patterns hold for $X_2$ and $X_3$. In Figure~\ref{fig:toy_uncertainty}B, we visualise the posterior distribution $p(\CRN\vert\data)$ using summary statistics of $(\idc(r\in\CRN))_{r\in\fullCRN}$, where $\idc(\cdot)$ denotes the indicator function. For each reaction~$r$, the posterior expectation of $\idc(r\in\CRN)$ is $\mathbb{P}(r\in\CRN\vert\data)$. We find that the reactions in $\CRN^\text{mode}$, namely reactions $13$, $18$, $30$, have the highest posterior probabilities, followed by reaction $1$, the ground-truth reaction omitted by $\CRN^\text{mode}$ (Figure~\ref{fig:toy_uncertainty}B, left panel). In Section~\ref{sec:consequences}, we gave the example that reactions $1$ ${(X_1 \rightarrow X_2)}$ and $30$ ${(X_1+X_3 \rightarrow X_2+X_3)}$ are practically indistinguishable. This is confirmed by the strong negative posterior correlation between $\idc(1\in\CRN)$ and $\idc(30\in\CRN)$ (Figure~\ref{fig:toy_uncertainty}B, right panel), suggesting that reactions $1$ and $30$ are alternatives that contribute similar explanations of the observed dynamics. The next strongest correlations are negative correlations for the pairs \splitatcommas{$(12,13), (13,15), (16,18), (17,18), (18,19), (18,20)$} and positive correlations for the pairs \splitatcommas{$(12,15), (16,19), (17,20)$}. These correlations suggest that the reaction sets \splitatcommas{$\{12,15\}, \{13\}$} are potential alternatives, and so are the reaction sets \splitatcommas{$\{16,19\}, \{17,20\}, \{18\}$}. In fact, these alternatives are elementary pairs of structurally indistinguishable reaction sets; see Supplementary Materials~S4.3 for the full list of elementary pairs. For instance, the dynamics induced by the reaction $X_3\xrightarrow{k_{13}}X_1+X_2$ are identical to the dynamics induced by the reactions $X_3\xrightarrow{k_{12}}X_2$ and $X_3\xrightarrow{k_{15}}X_1+X_3$ if $k_{12}=k_{13}=k_{15}$. Both sets of reactions, when parameterised this way, are equally likely to generate some given dataset. In other words, structurally indistinguishable CRNs cannot be discriminated by observed data alone, and thus should be accounted for when considering network uncertainty.

\begin{table}[t]
\newlength{\colwidth}
\settowidth{\colwidth}{\textbf{Approx.\ $L_0$}} 
\centering\small
\begin{tabular}{c *{4}{>{\centering\arraybackslash}m{\colwidth}}}
\toprule
\multirow[c]{2}{*}[-0.5ex]{CRN $R$} &
\multicolumn{4}{c}{Number of datasets (out of 25) where $\CRN\in\CRNset(\estset)$} \\
\cmidrule{2-5}
& $L_1$ & $\log L_1$ & Approx.\ $L_0$ & Horseshoe \\
\midrule
$\mathbf{\{1, 13, 18\}}$ & 21 & 25 & 25 & 25 \\
$\{1, 11, 14, 18\}$ & 9 & 24 & 25 & 24 \\
$\{1, 12, 15, 18\}$ & 3 & 23 & 19 & 24 \\
$\{1, 13, 16, 19\}$ & 7 & 25 & 25 & 25 \\
$\{1, 13, 17, 20\}$ & 3 & 20 & 20 & 22 \\
\bottomrule
\end{tabular}
\tblspace
\caption{Number of datasets where the ground-truth CRN (bold) and CRNs of four reactions that are structurally indistinguishable from it are included in the network ensemble $\CRNset(\estset)$ obtained with each penalty function. See Figure~\ref{fig:toy_uncertainty}C for reaction numbering.}\label{tbl:alt_crn}
\end{table}

\subsubsection{Representing network uncertainty}

To visually represent the composition of the 95\% HPD set, which we denote as $\CRNset^\text{HPD}$, and gain a sense of the variability of reactions within $\CRNset^\text{HPD}$, we introduce a hierarchical representation of $\CRNset^\text{HPD}$ (Figure~\ref{fig:toy_uncertainty}C). Recall that $\CRNset^\text{HPD}$ consists of CRNs that are capable of producing a reasonable fit to the data (Figure~\ref{fig:toy_uncertainty}A). The hierarchical representation of $\CRNset^\text{HPD}$ also helps with visualising alternative reaction sets, i.e. reaction sets that may be substituted for each other while preserving a reasonable fit to the data. The formal construction of this representation is described in Supplementary Materials~S2.4. Each node~$v$ of the hierarchy tree is characterised by a set of included reactions $\CRN^\text{inc}(v)$ and a set of excluded reactions $\CRN^\text{exc}(v)$. Let $\CRNset(v)$ be the set of CRNs in $\CRNset^\text{HPD}$ that include reactions $\CRN^\text{inc}(v)$ and exclude reactions $\CRN^\text{exc}(v)$. For the root node, we have $\CRN^\text{inc}(v)=\CRN^\text{exc}(v)=\varnothing$, so $\CRNset(v) = \CRNset^\text{HPD}$. A node~$v$ can have its CRN set $\CRNset(v)$ split into two and passed to its child nodes depending on the inclusion of some reaction~$r$. For the example in Figure~\ref{fig:toy_uncertainty}C, the root node split is determined by the inclusion of reaction $13$. At each node~$v$, reaction~$r$ is chosen to be the reaction with the highest posterior probability conditioned on the CRN set $\CRNset(v)$; we call this the reaction chosen at node $v$.

We terminate node splitting at node~$v$ if there is only one CRN in $\CRNset(v)$ that includes the reaction~$r$ chosen at node~$v$, or if $\CRN^\text{inc}(v) \in \CRNset(v)$. When the first termination criterion applies, the leaf node is labelled by the CRNs in $\CRNset(v)$, with reactions in $\CRN^\text{inc}(v)$ omitted. For example, the rightmost leaf node~$v$ ($\CRN^\text{inc}(v)=\{13\}$, $\CRN^\text{exc}(v)=\{18,16\}$) in Figure~\ref{fig:toy_uncertainty}C is associated with a CRN set $\CRNset(v)$ that consists of only one CRN, namely $\{13,17,20,30\}$. 
An example of the second termination criterion is the leftmost leaf node~$v$ ($\CRN^\text{inc}(v)=\{13,18,30\}$, $\CRN^\text{exc}(v)=\varnothing$), indicated by the absence of labelled reactions. This means that $\CRNset(v)$ includes the CRN $\{13,18,30\}$; the other 14 CRNs in $\CRNset(v)$ each consist of reactions $\{13,18,30\}$ and some extra reactions. 

The hierarchical representation corroborates the alternative reaction sets suggested by the posterior correlations. For instance, we previously suggested that $\{16,19\}, \{17,20\}, \{18\}$ are alternative reaction sets. In the subtree rooted at node~$v$ where $\CRN^\text{inc}(v)=\{13\}$, $\CRN^\text{exc}(v)=\{18\}$, we find that if reaction $16$ is included, then so is reaction $19$; otherwise, reactions $17$ and $20$ are included. We note that the dynamics induced by the reaction $X_3\xrightarrow{k_{13}}X_1+X_2$ are identical to the dynamics induced by the reactions $X_3\xrightarrow{k_{11}}X_1$ and $X_3\xrightarrow{k_{14}}X_2+X_3$ if $k_{11}=k_{13}=k_{14}$, yet reactions $11$ and $14$ do not appear in our hierarchical summary of the 95\% HPD set obtained with the $\log L_1$ penalty. In fact, for the $k_1=0.1,k_{18}=10$ dataset, the CRN $\{1,11,14,18\}$, which is structurally indistinguishable from $\CRN^\text{true}$, is included in the CRN ensemble $\CRNset(\estset)$ obtained with the approximate $L_0$ penalty, but is omitted by the other three penalty functions (not shown). This illustrates a limitation of our method---although our construction of $\CRNset(\estset)$ aims to include all CRNs with significant posterior probability, there is no guarantee that this will always occur. In Table~\ref{tbl:alt_crn}, we report the number of datasets where the ground-truth CRN and the CRNs that are structurally indistinguishable from it are included in the network ensemble $\CRNset(\estset)$ obtained with each penalty function. By far, the $L_1$ penalty results in the poorest coverage of these CRNs. Out of the nonconvex penalty functions, there is no clear winner.

The discrepancy of $\CRNset(\estset)$ across penalty functions is cause for concern. Recall that the normalisation constant in \eqref{eq:bayes_crn_sel} involves a summation over $\CRNset(\estset)$, which evidently can be different for each penalty function used. On the other hand, the exact counterpart \eqref{eq:bayes_crn} exhaustively sums over all possible CRNs. If a CRN $R$ whose unnormalised posterior ${p(\CRN) \exp (-\textsc{bic}(\CRN) /2 )}$ contributes significantly to the exact normalisation constant is omitted by $\CRNset(\estset)$, then \eqref{eq:bayes_crn_sel} will underestimate the normalisation constant, potentially distorting the obtained posterior summaries. To be clear, there is a single true posterior \eqref{eq:bayes_crn} that does not depend on the penalty function, but our approximate posterior \eqref{eq:bayes_crn_sel} may vary depending on the penalty function used. In Figure~S3, we find that for some datasets, the CRN ensemble $\CRNset(\estset)$ obtained with the $L_1$ penalty omits many CRNs with relatively large unnormalised posterior values. As for the nonconvex penalty functions, the variation across the CRN ensembles $\CRNset(\estset)$ is less than the variation across the CRN ensembles $\CRNset_\text{base}(\estset)$ obtained without the recombination stage. We will revisit the impact of the recombination stage in our case studies that follow.

\subsection{Case study 1: \texorpdfstring{\textalpha}{α}-pinene isomerisation}\label{sec:pinene}

\begin{figure}
\includegraphics[width = 0.99\textwidth]{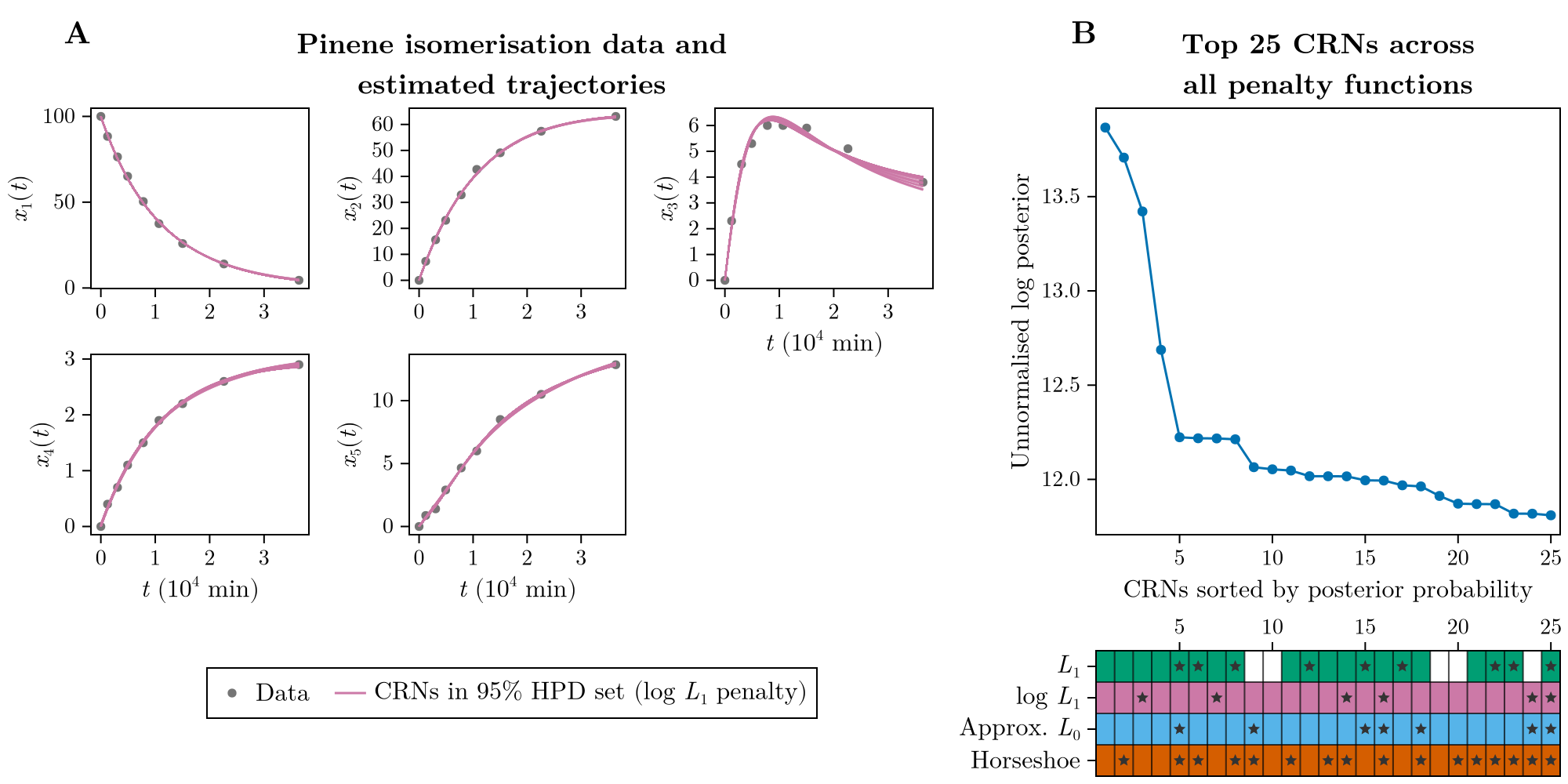}
\caption{(A) Experimental data for the \textalpha-pinene isomerisation case study along with trajectories reconstructed using the MLEs of the CRNs in the 95\% HPD set obtained with the $\log L_1$ penalty. (B) Unnormalised log posterior of the top 25 CRNs pooled from the CRN ensembles $\CRNset(\estset)$ obtained with each penalty function for the \textalpha-pinene isomerisation case study. Shaded cells indicate whether a CRN is present in the CRN ensemble $\CRNset(\estset)$ obtained with some penalty function. Starred cells correspond to CRNs in $\CRNset(\estset)$ (after recombination) but not in $\CRNset_\text{base}(\estset)$ (before recombination).}
\label{fig:pinene}
\end{figure}

In real-world applications, we expect network uncertainty to be more pronounced as most datasets do not sample observations as densely as our simulation study in Section~\ref{sec:toy}. \textcite{fuguitt_liquid_1945} obtained experimental data for the thermal isomerisation of \textalpha-pinene at 9 time points, which is a reaction system consisting of \textalpha-pinene ($X_1$), dipentene ($X_2$), allo-ocimene ($X_3$), \textalpha- and \textbeta-pyronenes ($X_4$), and a dimer ($X_5$). This 80-year-old dataset has been extensively used to benchmark ODE parameter inference methods \cite{stewart_bayesian_1981,rodriguez-fernandez_novel_2006,miro_deterministic_2012,brunel_tracking_2015,rakhshani_hierarchy_2016}, which assume a fixed CRN structure. \textcite{fuguitt_liquid_1945} proposed the CRN
\begin{equation}
    X_1 \rightarrow X_2, \quad X_1 \rightarrow X_3 \rightarrow X_4, \quad 2X_3 \rightleftharpoons X_5.
\end{equation}
This CRN was later extended by \textcite{stewart_bayesian_1981} to include the reactions ${X_4 \rightarrow X_3}$ and ${2X_1 \rightarrow X_5}$, who concluded that the extended model should be preferred. To our knowledge, the only work that performs inference over a combinatorial space of reactions for this dataset is \textcite{tsu_computational_2019}, where a family of CRNs satisfying some simplifying assumptions were enumerated exhaustively. These assumptions were: (i) $X_1$ does not participate as a product, (ii) $X_2$ and $X_5$ do not participate as reactants, and (iii) all reactions are unimolecular, except that dimerisation and dissociation of $X_5$ involves two molecules of another species. Moreover, $X_4$ was disregarded during inference in \textcite{tsu_computational_2019} as the amount of $X_4$ was not directly measured due to experimental limitations as reported in~\textcite{fuguitt_liquid_1945}. Instead, the amount of $X_4$ at any given time is assumed to be 3\% of the reacted amount of $X_1$, an assumption which \textcite{fuguitt_liquid_1945} make use of to produce the data that we apply our method to.

\begin{table}[t]
\centering\tablefontsize
\begin{tabular}{*{9}{c}}
\toprule
\multirowcell{2}{\small Reaction} &
\multicolumn{4}{c}{\small Appears in} &
\multicolumn{4}{c}{\small Posterior reaction probability} \\
\cmidrule{2-9}
& \makecell{Original\\CRN} & \makecell{Extended\\CRN} & \makecell{5-reaction\\CRN} & \makecell{6-reaction\\CRN} & \makecell{$L_1$\\penalty} & \makecell{log $L_1$\\penalty} & \makecell{Approx.\\$L_0$ penalty} & \makecell{Horseshoe\\penalty}
\\
\midrule
$X_1 \rightarrow X_2$ & \checkmark & \checkmark & \checkmark & \checkmark & 0.8005 & 0.8749 & 0.7449 \\
$X_1 \rightarrow X_3$ & \checkmark & \checkmark & \checkmark & \checkmark & 1.0000 & 1.0000 & 1.0000 \\
$X_3 \rightarrow X_4$ & \checkmark & \checkmark & NA & NA & 0.2134 & 0.1647 & 0.0361 \\
$2X_3 \rightarrow X_5$ & \checkmark & \checkmark & \checkmark & \checkmark & 0.9999 & 0.9999 & 1.0000 \\
$X_5 \rightarrow 2X_3$ & \checkmark & \checkmark & \checkmark & & 0.5466 & 0.5411 & 0.4164 \\
$X_4 \rightarrow X_3$ & & \checkmark & NA & NA & 0.4998 & 0.3863 & 0.3116 \\
$2X_1 \rightarrow X_5$ & & \checkmark & \checkmark & \checkmark & 0.9987 & 0.9987 & 1.0000 \\
$X_2 \rightarrow X_3$ & & & & \checkmark & 0.5602 & 0.6392 & 0.6333 \\
$X_5 \rightarrow 2X_2$ & & & & \checkmark & 0.3361 & 0.3169 & 0.3444 \\
\bottomrule
\end{tabular}
\tblspace
\caption{Posterior probability of reactions featured in literature for the \textalpha-pinene isomerisation case study. The original CRN is taken from \textcite{fuguitt_liquid_1945}, the extended CRN is taken from \textcite{stewart_bayesian_1981}, the 5- and 6-reaction CRN are taken from \textcite{tsu_computational_2019}.  Reactions involving $X_4$ are marked as NA for \textcite{tsu_computational_2019} as the authors excluded $X_4$ from their analysis.}\label{tbl:pinene_post}
\end{table}

For our application, we use assumption (iii) alone to construct a library of 20 reactions over all 5 species (including $X_4$); see Supplementary Materials~S3.2. The optimisation details described in Section~\ref{sec:toy} apply here. We find that for all penalty functions, the 95\% HPD set consists of more than 100 CRNs, indicating a large degree of network uncertainty. The trajectories reconstructed using the MLEs of the CRNs in the 95\% HPD set obtained with the $\log L_1$ penalty match the data well (Figure~\ref{fig:pinene}A). Out of the 5 species, $X_3$ shows the most variation over its trajectories, which suggests that network uncertainty may be narrowed down by collecting more data for $X_3$. To assess whether our inferred CRNs are in line with reactions proposed in literature, we report in Table~\ref{tbl:pinene_post} the posterior probability of each reaction featured in the literature. This includes reactions reported from the exhaustive approach of \textcite{tsu_computational_2019}. With the exception of $X_3\rightarrow X_4$, all other literature-featured reactions have an appreciable posterior probability. Upon inspecting the hierarchical representation of the 95\% HPD set (Figure~S5), we find that the exception of $X_3\rightarrow X_4$ can be explained by the fact that for most of the posterior mass, the production of $X_4$ is accounted for by including the reaction $X_1\rightarrow X_4$. This comes as no surprise because the observations of $X_4$ were set as a constant proportion of the reacted $X_1$ \cite{fuguitt_liquid_1945}. A further analysis of the hierarchical representation can be found in Supplementary Materials~S5.2. Notably, the analysis reveals higher-order ambiguities that are not immediately obvious from inspecting the posterior correlations.

Again, we note that each penalty function produces a different CRN ensemble $\CRNset(\estset)$, resulting in slight variations in reaction probabilities across penalty functions (Table~\ref{tbl:pinene_post}). Nevertheless, the unnormalised posterior of a CRN $R$, ${p(\CRN) \exp (-\textsc{bic}(\CRN) /2 )}$, does not depend on the penalty function, allowing us to rank the CRNs pooled over all CRN ensembles by their unnormalised posterior (Figure~\ref{fig:pinene}B). Out of all penalty functions, the $L_1$ penalty has the poorest coverage of the top CRNs. The top 25 CRNs are found by all nonconvex penalty functions. This concordance is corroborated by visually similar posterior correlations obtained with nonconvex penalty functions (Figure~S4). Thus, the variation of $\CRNset(\estset)$ across nonconvex penalty functions only concern CRNs with lower posterior probabilities. We note that the recombination procedure is instrumental in finding the top CRNs. Without the recombination procedure, none of the penalty functions are able to individually find all top 8 CRNs. 

In addition, we have enumerated elementary pairs for this case study and the next (Section~\ref{sec:pyridine}), but the results are not of enough interest to warrant discussion here; we report these results in Supplementary Materials~S4.3 for completeness.

\subsection{Case study 2: pyridine denitrogenation}\label{sec:pyridine}

In this case study, we investigate a scenario where, due to a much larger model space, there is insufficient signal in the data to provide confident insight into the CRN structure. The reaction system under investigation here features the denitrogenation of pyridine \cite{bock_numerical_1981}, with data reported in \textcite{schittkowski_collection_2009}. This system features 7 species, and data are reported for 6 species at 12 time points. The unreported species (pentane) is assumed to not participate as a reactant, so we proceed without modelling its dynamics. A CRN consisting of 11 reactions has been proposed for this system in \textcite{bock_numerical_1981}. We estimate rate constants for this CRN by maximum likelihood estimation, and consider this a gold-standard solution to compare to later. We perform CRN inference with a library of 67 candidate reactions, which includes the 11 reactions of the gold-standard CRN. Since this library is much larger than our previous libraries, we increase the number of optimisation starting points to $N_\text{start}=256$; other optimisation details are unchanged. See Supplementary Materials~S3.3 for details about the candidate reactions and the gold-standard CRN.

\begin{figure}
\includegraphics[width = 0.99\textwidth]{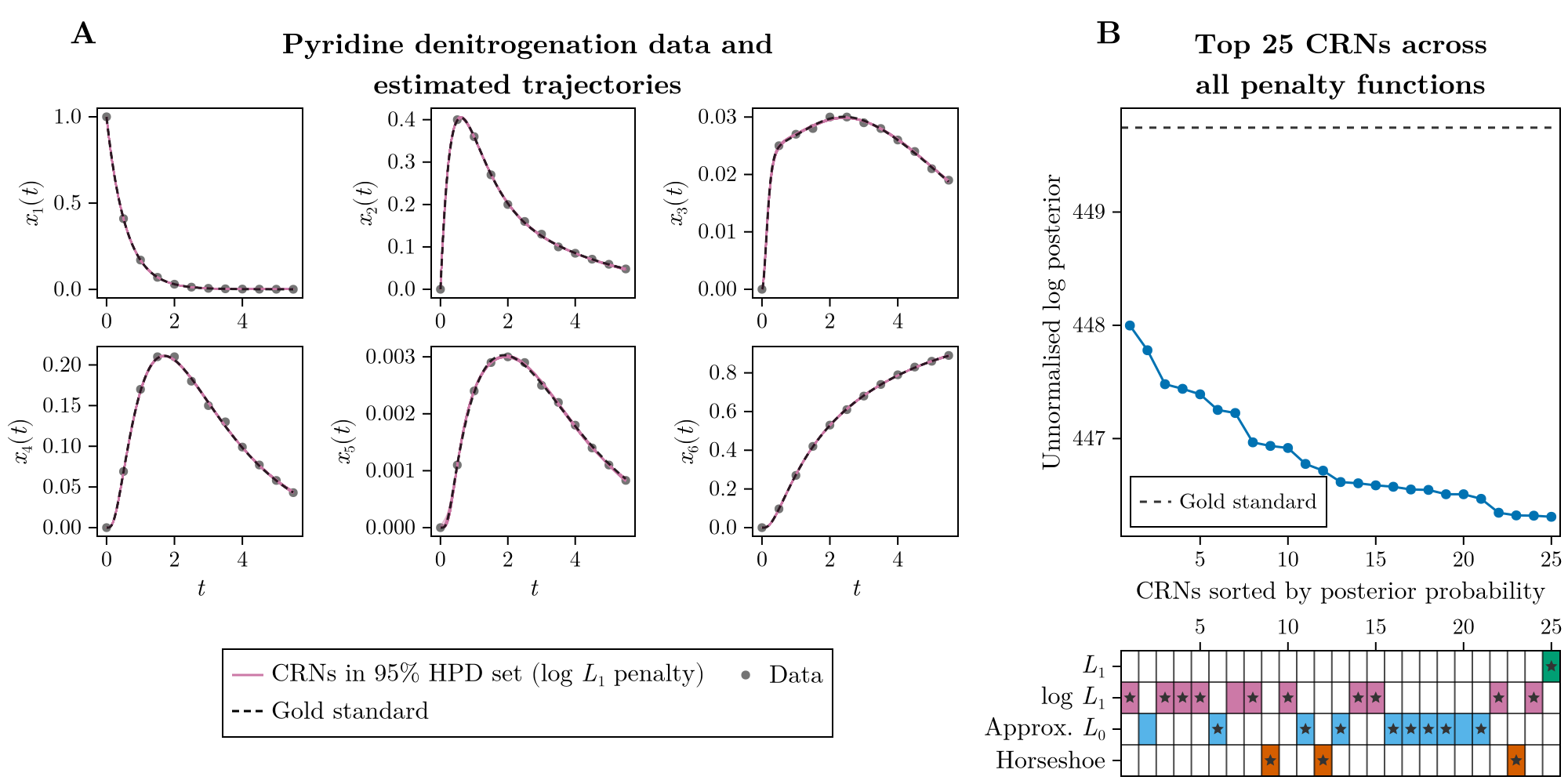}
\caption{(A) Experimental data for the pyridine denitrogenation case study along with trajectories reconstructed using the MLEs of the gold-standard CRN and CRNs in the 95\% HPD set obtained with the $\log L_1$ penalty. (B) Unnormalised log posterior of the top 25 CRNs pooled from the CRN ensembles $\CRNset(\estset)$ obtained with each penalty function for the pyridine denitrogenation case study. The dashed line corresponds to the unnormalised log posterior of the gold-standard CRN \cite{bock_numerical_1981}. Shaded cells indicate whether a CRN is present in the CRN ensemble $\CRNset(\estset)$ obtained with some penalty function. Starred cells correspond to CRNs in $\CRNset(\estset)$ (after recombination) but not in $\CRNset_\text{base}(\estset)$ (before recombination).}
\label{fig:pyridine}
\end{figure}

CRNs in the 95\% HPD set obtained with the $\log L_1$ penalty provide model fits that are visually as good as the gold-standard fit (Figure~\ref{fig:pyridine}A). With few data points to constrain the vast model space with, we expect network uncertainty to be greatly pronounced. Yet, the 95\% HPD sets for this case study (consisting of $4, 78, 36, 21$ CRNs for the four penalty functions, respectively) are smaller than those of the previous case study ($>\!100$ CRNs for all penalty functions). This is due to a failure to find all CRNs with a high true posterior probability, evident from the fact that the gold-standard CRN was not recovered by any CRN ensemble $\CRNset(\estset)$, despite having a larger unnormalised posterior value than all CRNs in the CRN ensembles found (Figure~\ref{fig:pyridine}B). This poor truncation of the posterior results in a severe underestimation of the normalisation constant in \eqref{eq:bayes_crn}. Subsequently, the approximate posterior probabilities~\eqref{eq:bayes_crn_sel} of the CRNs found are grossly overestimated relative to their exact counterparts, leading to 95\% HPD sets that in truth capture less than 95\% of the exact posterior. If a gold standard were not available, the issue of poor truncation could be identified by checking for overlap between the CRN ensembles $\CRNset(\estset)$ obtained with each penalty function among the CRNs with the highest unnormalised posterior. In Figure~\ref{fig:pyridine}B, we find that each of the top 25 CRNs (ranked by unnormalised posterior) appears in only one CRN ensemble. The lack of an overlap suggests that there are CRNs with relatively large unnormalised posterior values that have not been found. Moreover, the posterior correlations obtained with each penalty function are dissimilar (Figure~S6).

We emphasise that although the approximate posterior probabilities cannot be trusted in an absolute sense, relative comparisons of the unnormalised posterior remain valid. None of the top 25 CRNs are found with the $L_1$ penalty, illustrating again its poor ability to find parsimonious CRNs. On the other hand, the CRNs found with the $\log L_1$ penalty contribute the most posterior mass. In Supplementary Materials~S5.3, we show that the gold-standard CRN is partially recovered by a hierarchical representation of the 95\% HPD set obtained with the $\log L_1$ penalty (noting that the true coverage of this 95\% HPD set is less than 95\%), and explain why some gold-standard reactions are missed. Out of the 11 gold-standard reactions, 5 reactions are included by every CRN of the 95\% HPD set, and all reactions but one are featured in the hierarchical representation.

\afterpage{\FloatBarrier}

\section{Discussion}

In this work, we have adapted sparse regularisation methods to perform Bayesian model selection without iteratively sampling model structures. Typical implementations of sparse regularisation often return a single network structure. We find that ignoring network uncertainty leads to unreliable predictions (Figure~\ref{fig:toy_mode_traj}), whereas accounting for network uncertainty helps identify alternative reactions of a system (Figure~\ref{fig:toy_uncertainty} and Table~\ref{tbl:pinene_post}). In Section~\ref{sec:toy}, we demonstrated how network uncertainty can arise due to structural indistinguishability---the ability for distinct CRNs to produce identical dynamics, or due to practical indistinguishability---the lack of information in observed data to discriminate between distinct CRNs of similar dynamics. To our knowledge, the tree representation of a 95\% HPD set of CRNs (e.g., Figure~\ref{fig:toy_uncertainty}C) is a novel visualisation of network uncertainty for CRN inference, allowing network ambiguities to be highlighted. In the field of \review{equation learning}, there are recent developments in uncertainty quantification \cite{hirsh_sparsifying_2022,fasel_ensemble-sindy_2022,sun_bayesian_2022,niven_dynamical_2024}. However, in these works, model uncertainty, if addressed at all, is only reported at a univariate level, such as inclusion probabilities of library terms. Uncertainty quantification in \review{equation learning} can readily benefit from the higher-order descriptions of network uncertainty (alternative reaction sets, posterior correlations, tree visualisations) illustrated in this work. 

Sparse regularisation methods implemented with multi-start optimisation can be naturally extended to investigate network uncertainty, but this is rarely done. We have quantified network uncertainty via an approximate posterior distribution~\eqref{eq:bayes_crn_sel} over CRNs which correspond to local optima found in parameter space, while introducing minimal implementation overhead. A limitation is that our posterior results may be sensitive to the quality and diversity of the estimates found during optimisation. The recombination stage is instrumental in alleviating this issue, as it discovers CRNs with high posterior probabilities missed by continuous optimisation (Figures~\ref{fig:pinene} and S3). We believe that the construction of elementary pairs of structurally indistinguishable reaction sets (Section~\ref{sec:elementary}) can potentially make the recombination stage more effective. However, in this work, we have used these elementary pairs as an external source of ground truth to evaluate how well each penalty function recovers structurally indistinguishable CRNs (Table~\ref{tbl:alt_crn}). As such, the use of elementary pairs for recombination is considered future work. We note that this theory on elementary pairs applies only to reactions governed by mass-action kinetics. On the other hand, the inference procedure can be generalised to incorporate other reaction kinetics. For example, the use of Michaelis-Menten kinetics would introduce saturation parameters~\cite{segelModelingDynamic1984}, which could be optionally penalised depending on the modelling assumptions.

We find that the $L_1$ penalty function results in less parsimonious CRNs compared to the nonconvex penalty functions. When cast as a sparse regularisation problem, CRN inference resembles the problem of variable selection in statistical literature. The $L_1$ (or lasso) penalty is a popular choice for inducing sparsity, yet we have demonstrated that its performance for CRN inference is poor. The poor performance cannot be salvaged by simply increasing the penalty hyperparameter, as the largest penalty hyperparameters we used in this work for the $L_1$ penalty resulted in overly simple CRNs during the pruning stage (results not shown). It is unclear whether this failure of the $L_1$ penalty is because the optima of the resulting loss function do not correspond to parsimonious CRNs, or because the basins of attraction for the optima corresponding to the most parsimonious reaction networks are too narrow. From a Bayesian perspective, the $L_1$ penalty corresponds to an exponential prior on the rate constants, whereas the nonconvex penalty functions correspond to heavy-tail prior distributions. The nonconvex penalty functions better reflect our prior belief that most candidate reactions are not present in the system.

Our posterior distribution~\eqref{eq:bayes_crn_sel} is approximate in two senses, namely due to approximating the model evidence \eqref{eq:bic} and truncating the exact posterior~\eqref{eq:bayes_crn}. The BIC approximation of the model evidence assumes that the number of observations $\lvert\data\rvert$ is asymptotically large, which is violated in our case studies. The BIC approximation has an $\bigO(\log\lvert\data\rvert)$ error, and ignores any prior placed on parameters and the curvature of the log-likelihood in parameter space. A more accurate alternative is to use a Laplace approximation of the model evidence \cite{llorente_marginal_2023}, which is based on a second-order Taylor approximation of the log-likelihood. This work can benefit from accurate model evidence computation methods that can scale up to thousands of CRNs. The validity of our approximate posterior~\eqref{eq:bayes_crn_sel} also relies on $\CRNset(\estset)$ covering the bulk of the exact posterior~\eqref{eq:bayes_crn}. This is influenced by the design of the loss function~\eqref{eq:loss_func}, how it is optimised, and how parameter estimates~$\estset$ are mapped to a CRN ensemble~$\CRNset(\estset)$. When there is insufficient data to constrain a high-dimensional model space, there are too many CRNs with non-negligible posterior mass to be found with multi-start local optimisation. It is unclear whether mapping all local optima of a class of loss functions to CRN space will cover all relevant CRNs. A potential improvement is to use the CRN prior~\eqref{eq:uniform_prior} to inform the design of the penalty function. But even if the optima of a prior-informed loss function were to cover all relevant CRNs, some local optima may not have large enough basins of attraction for them to be practically found within reasonable computation. Our recombination procedure is a partial answer to these concerns, which we have shown to improve the concordance between CRN ensembles $\CRNset(\estset)$ obtained with nonconvex penalty functions. We also suspect that local optima can be more robustly found by combining global and local optimisation strategies, an approach advocated by \textcite{villaverde_benchmarking_2019} in their benchmarking work on fitting large kinetic models. Multimodal search strategies that aim to promote the diversity of the solutions throughout the search process should be considered; see~\textcite{das_real-parameter_2011} for a review.

Understanding network uncertainty is important in biological systems beyond the biochemical settings considered in this work. Our departure from relying on a single network structure, which is often the case for sparse regularisation methods, allows for network uncertainty to be quantified as a posterior distribution over networks. Reaction ambiguities elicited from a network posterior can guide how a biological system should be perturbed in future experiments to optimise their utility in narrowing down network uncertainty. For instance, given a handful of networks that plausibly explain some previous data, one can optimise the initial conditions of a future experiment to maximise the information available to discriminate between the given networks \cite{blanquero_global_2016,silk_model_2014}. There is a need to develop methods that translate descriptions of network uncertainty to optimal experimental design. The cycle of experimental design and network uncertainty quantification is a promising paradigm for catalysing scientific discovery in biology.

\vspace{2ex}
\subsection*{Acknowledgements}

We thank the mathematical research institute MATRIX in Australia where part of this research was performed, where we had engaging discussions with Alejandro Villaverde, Jae Kyoung Kim, and Ziwen Zhong. We also thank Alexander Browning, Ahmad Mokhtar, and anonymous reviewers for helpful comments on the manuscript.

\subsection*{Data availability}

The code and data used in this work are available at \url{https://github.com/ysfoo/crn-inference}.

\AtNextBibliography{\small}
\emergencystretch 2em
\printbibliography[heading=bibintoc]

\end{refsection}

\pagebreak
\begin{center}
\Large{Supplementary Materials: Quantifying network uncertainty in chemical reaction network inference}
\end{center}
\setcounter{section}{0}
\setcounter{equation}{0}
\setcounter{figure}{0}
\setcounter{table}{0}
\setcounter{algorithm}{0}
\setcounter{theorem}{0}
\makeatletter
\renewcommand{\thesection}{S\arabic{section}}
\renewcommand{\theequation}{S\arabic{equation}}
\renewcommand{\thefigure}{S\arabic{figure}}
\renewcommand{\thetable}{S\arabic{table}}
\renewcommand{\thealgorithm}{S\arabic{algorithm}}
\renewcommand{\thetheorem}{S\arabic{theorem}}

\begin{refsection}

\section{Optimisation details}

\label{sec:optim}

\subsection{Objective function with transformed variables}

In Section~\ref{sec:param_infer} of the main text, we noted that for each reaction $r\in\fullCRN$, the dimensionless parameter $k_r/\kappa_r$ is penalised, instead of the original rate constant $k_r$. Furthermore, we perform optimisation on the log-transformed variables. Let $\eta_r = \log(k_r/\kappa_r)$ for each $r\in\fullCRN$ and $\varsigma_s = \log \sigma_s^2$ for each $s=1,\ldots,S$. Formally, we minimise the loss function
\begin{align}\label{eq:transformed_func}
    l(\bm\eta,\bm\varsigma; \lambda) 
    &= -\log p(\data\vert \exp(\bm\eta) \odot \bm\kappa, \exp(\bm\varsigma) ) + \sum_{r \in \fullCRN} \text{pen}(\exp(\eta_r); \lambda) \\
    &= -\log p(\data\vert \mathbf{k}, \bm\sigma^2) + \sum_{r \in \fullCRN} \text{pen}(k_r/\kappa_r; \lambda)  \nonumber
\end{align}
over $\bm\eta$ and $\bm\varsigma$, where $\bm\eta=(\eta_r)_{r\in\fullCRN}$, $\bm\kappa=(\kappa_r)_{r\in\fullCRN}$, $\bm\varsigma=(\varsigma_s)_{s=1,\ldots,S}$, $\odot$ denotes elementwise multiplication, and $\exp(\cdot)$ is performed elementwise.

\subsection{Scaling factors for reaction rate constants}

Recall that the ODE system is given by
\begin{equation*}
    x'_s = \sum_{r\in\fullCRN} (m_{rs}^+ - m_{rs}^-) k_r \prod_{s'=1}^S x_{s'}^{m_{rs'}^-}
\end{equation*}
for each $s=1,\ldots,S$. To determine the scaling factors $\bm\kappa$, that is, to estimate the order of magnitude of each $k_r$, we need estimates of $\mathbf{x}$ and $\mathbf{x}'$ that are not tied to any specific CRN. To this end, let $\hat{\mathbf{x}}$ and $\hat{\mathbf{x}}'$ denote estimates of $\mathbf{x}$ and $\mathbf{x}'$ obtained by fitting smoothing B-splines to the observed trajectories $\data$~\cite{stickel_data_2010}.

For each reaction $r\in\fullCRN$, we set $\kappa_r$ to be
\begin{equation}\label{eq:estim_kappa}
    \max_{s\in \mathcal{S}(r)} \frac{1}{\lvert m_{rs}^+ - m_{rs}^- \rvert}\frac{\max_{t\in[0,t_N]} \hat{x}'_s(t)  - \min_{t\in[0,t_N]} \hat{x}'_s(t) }{\max_{t\in[0,t_N]} \prod_{s'=1}^S \hat{x}_{s'}(t)^{m_{rs'}^-} - \min_{t\in[0,t_N]} \prod_{s'=1}^S \hat{x}_{s'}(t)^{m_{rs'}^-}},
\end{equation}
where 
\begin{equation*}
    \mathcal{S}(r) \coloneqq \{ s=1,\ldots,S \colon m_{rs}^+ - m_{rs}^-\neq 0 \}
\end{equation*}
denotes the set of species whose dynamics are directly affected by reaction $r$. The rationale behind \eqref{eq:estim_kappa} is that for each reaction $r$ present in the system, we assume that there is some species $s$ such that the dynamics of species $s$ are dominated by reaction $r$ alone. Under this assumption, we expect $k_r$ to have a similar order of magnitude to the ratio of the range of $dx_s/dt$ to the range of $(m_{rs}^+ - m_{rs}^-) \prod_{s'=1}^S x_{s'}^{m_{rs'}^-}$. As this assumption does not necessarily hold, $\kappa_r$ is only a crude estimate for the order of magnitude of $k_r$. The (outer) maximum in \eqref{eq:estim_kappa} implies that $\kappa_r$ behaves more like an upper bound, such that $k_r/\kappa_r$ is unlikely to be much larger than 1. We confirmed this empirically with our simulation study in Section~3.1, which features ground-truth rate constants of different scales. Out of all estimated values of $k_r/\kappa_r$ across all 95\% HPD sets, only 0.37\% of them were larger than $10^1$, while 7.1\% of them were less than $10^{-3}$. This suggests that our scaling factors are adequate.

\subsection{Starting points and bounds for optimisation}\label{sec:opt_init_bound}

When minimising \eqref{eq:transformed_func}, we set the starting values for $\bm\eta$ and $\bm\varsigma$ with two separate procedures. We choose $N_\text{start}$ starting values for $\bm\eta$ using quasi-Monte Carlo sampling, which samples the parameter space more `evenly', within the interval $[\log 10^{-4}, \log 10^0]$. Specifically, we use Sobol sequences randomised with the Owen scramble~\cite{owen_practical_2023}. On the other hand, the starting values for $\bm\varsigma$ are fixed based on the smoothing spline estimates $\hat{\mathbf{x}}$. For each $s=1,\ldots S$, the starting value of $\varsigma_s$ is set as
\begin{equation}\label{}
    \varsigma^\text{init}_s = \log \!\left( \frac{1}{N}\sum_{n=1}^N (y_{ns}-\hat{x}_s(t_n))^2 \right).
\end{equation}
We impose optimisation lower bounds for all parameters. For the rate constants, we impose $k_r \ge \epsilon = 10^{-10}$ for each reaction $r\in\fullCRN$ as some penalty functions diverge as $k_r\rightarrow 0^+$. We impose lower bounds for noise variances to prevent overfitting to some species' trajectory at the expense of the fits for the other species. For our case studies, the lower bound for the noise standard deviation $\sigma_s$ is set as the reporting precision, which is 0.05 for the $\alpha$-pinene isomerisation case study and $10^{-5}$ for the pyridine denitrogenation case study. For the simulation study, we impose ${\sigma_s \ge 0.01 \sqrt{\exp(\varsigma^\text{init}_s)}}$.

\subsection{Penalty hyperparameter values}

In this work, we minimise the function \eqref{eq:transformed_func} for $N_\text{hyp}=10$ values of the penalty hyperparameter $\lambda$, which we denote as $\lambda_1 < \cdots < \lambda_{10}$. The sequence of hyperparameter values are hand-picked for each penalty function $\text{pen}(k_r/\kappa_r; \lambda)$. Each sequence of hyperparameter values is geometric, so that a few orders of magnitude of $\lambda$ are covered. Recall that when minimising \eqref{eq:transformed_func}, we impose a lower bound of $\varepsilon=10^{-10}$ on $k_r/\kappa_r$ for each $r\in\fullCRN$. Thus, if a reaction $r$ does not contribute to the system dynamics significantly, its corresponding penalty should be $\text{pen}(\varepsilon; \lambda)$. Meanwhile, if a reaction $r$ contributes to the system dynamics significantly, we expect its corresponding penalty to be roughly $\text{pen}(1; \lambda)$. The difference ${\text{pen}(1; \lambda)-\text{pen}(\varepsilon; \lambda)}$ is thus a proxy for how much the negative log-likelihood is penalised by each reaction (that contributes to the dynamics significantly), which we use to calibrate our choice of hyperparameter values.

For the approximate $L_0$ penalty
\begin{equation}\label{eq:approxL0}
    \text{pen}(k_r/\kappa_r; \lambda) = \lambda (k_r/\kappa_r)^{0.1},
\end{equation}
we have $\text{pen}(1; \lambda)-\text{pen}(\varepsilon; \lambda) = 0.9\lambda$. Recalling that the Bayesian information criterion (BIC) penalises the negative log-likelihood by $(\log\lvert\data\rvert)/2$ for each model parameter, we centre our sequence of hyperparameter values around a typical value of $(\log\lvert\data\rvert)/2$, say $(\log 100)/2 = 2.3$. We choose $\lambda_1,\ldots,\lambda_{10}=2^{-3},\ldots,2^6$ as the sequence of hyperparameter values for the approximate $L_0$ penalty.

For the log-scale $L_1$ penalty
\begin{equation}\label{eq:logL1}
    \text{pen}(k_r/\kappa_r; \lambda) = \lambda \lvert\log(k_r/\kappa_r) - \log \epsilon \rvert,
\end{equation}
we have $\text{pen}(1; \lambda)-\text{pen}(\varepsilon; \lambda) = \log(10^{10})\lambda$. We use the previous sequence of hyperparameter values divided by $\log(10^{10})$, namely $\lambda_1,\ldots,\lambda_{10}=2^{-3}/\log(10^{10}),\ldots,2^6/\log(10^{10})$.

For the $L_1$ penalty
\begin{equation}\label{eq:L1}
    \text{pen}(k_r/\kappa_r; \lambda) = \lambda (k_r/\kappa_r),
\end{equation}
we have $\text{pen}(1; \lambda)-\text{pen}(\varepsilon; \lambda) = \lambda$. Initially, we used the sequence of hyperparameter values $\lambda_1,\ldots,\lambda_{10}=2^{-3},\ldots,2^6$, but found that this choice often did not provide sufficient penalty strength. We suspect this occurred as $x\mapsto x^{0.1}$ from \eqref{eq:approxL0} and $x\mapsto\log x$ from \eqref{eq:logL1} increase more rapidly around $x=\varepsilon$ than $x=1$, but the same is not true for $x\mapsto x$ from \eqref{eq:L1}. Thus, we use a wider range of hyperparameter values $\lambda_1,\ldots,\lambda_{10}=4^{-3},\ldots,4^6$ for the $L_1$ penalty.

For the horseshoe-like penalty 
\begin{equation}\label{eq:hslike}
    \text{pen}(k_r/\kappa_r; \lambda) = -\log(\log (1+1/(\lambda k_r/\kappa_r)^2)),
\end{equation}
we do not reuse the same approach since this penalty does not scale linearly with $\lambda$. As $\lambda\rightarrow\infty$, we have
\begin{align*}
    \text{pen}(1; \lambda)-\text{pen}(\varepsilon; \lambda)
    &= - \log ( \log (1 + \lambda^{-2}) ) + \log ( \log (1 + \lambda^{-2}\varepsilon^{-2}) ) \\
    &= - \log ( \lambda^{-2} + \bigO(\lambda^{-4}) ) + \log ( \lambda^{-2}\varepsilon^{-2} + \bigO(\lambda^{-4}) ) \\
    &= - \log ( \lambda^{-2} ) + \log ( \lambda^{-2}\varepsilon^{-2} ) + \bigO(\lambda^{-2}) \\
    &= -2 \log ( \varepsilon) + \bigO(\lambda^{-2}). 
\end{align*}
In other words, $\lim_{\lambda\rightarrow\infty}(\text{pen}(1; \lambda)-\text{pen}(\varepsilon; \lambda)) = -2\log \varepsilon$. Meanwhile, as $\lambda\rightarrow 0^+$, we have
\begin{align*}
    \text{pen}(1; \lambda)-\text{pen}(\varepsilon; \lambda)
    &= \log \frac{\log (1 + \lambda^{-2}\varepsilon^{-2})}{\log (1 + \lambda^{-2})} \\
    &= \log \frac{\log (\lambda^{-2}\varepsilon^{-2}) + \bigO(\lambda^2) }{\log (\lambda^{-2}) + \bigO(\lambda^2)} \\
    &= \log \frac{-2\log\lambda-2\log\varepsilon + \bigO(\lambda^2) }{-2\log\lambda + \bigO(\lambda^2)} \\
    &= \log \frac{1 + \frac{\log\varepsilon}{\log\lambda} + \bigO(\lambda^2/\log\lambda) }{1 + \bigO(\lambda^2/\log\lambda)} \\
    &= \log \!\left( 1 + \frac{\log\varepsilon}{\log\lambda} + \bigO(\lambda^2/\log\lambda) \right) \\
    &= \frac{\log\varepsilon}{\log\lambda} + \bigO(1/(\log\lambda)^2).
\end{align*}
This implies that if we want $\text{pen}(1; \lambda)-\text{pen}(\varepsilon; \lambda) \ll 1$, then $\lambda$ would need to be extremely small, which we avoid in case of numerical issues. We choose a geometric sequence $\lambda_1,\ldots,\lambda_{10}$ such that $\text{pen}(1; \lambda_1)-\text{pen}(\varepsilon; \lambda_1)$ is close to 1 and $\text{pen}(1; \lambda_{10})-\text{pen}(\varepsilon; \lambda_{10})$ is close to $-2\log \varepsilon = 46.1$. Specifically, we choose $\lambda_1,\ldots,\lambda_{10}=32^{-3},\ldots,32^6$, noting that $\text{pen}(1; \lambda_1)-\text{pen}(\varepsilon; \lambda_1)=1.2$ and $\text{pen}(1; \lambda_{10})-\text{pen}(\varepsilon; \lambda_{10})=43.1$.

\clearpage

\section{Posterior computation details}

We provide a summary of our method in Algorithm~\ref{alg:inference}, with some procedures to be elaborated in the following subsections.

\algrenewcommand\algorithmicrequire{\textbf{Input:}}

\begin{algorithm}
\caption{Bayesian inference over CRN structures obtained with sparse regularisation}\label{alg:inference}
\begin{algorithmic}[1]
\Require \begin{minipage}[t]{0.95\linewidth}
Library of candidate reactions $\fullCRN$, time-series data $\data$, hyperparameter values $\lambda_1, \ldots, \lambda_{N_\text{hyp}}$, penalty function $\text{pen}(\cdot; \lambda)$, CRN prior $p(\CRN)$
\end{minipage}
\State Determine scaling factors for rate constants \Comment{Section~S1.2}
\State $\estset \gets$ empty set of parameter estimates
\For{$\lambda\in\{\lambda_1, \ldots, \lambda_{N_\text{hyp}}\}$}
    \State Choose $N_\text{start}$ randomised starting points for optimisation \Comment{Section~S1.3}
    \For{$i=1$ to $N_\text{start}$}        
        \State $\est \gets \argmin_{\bm\theta} l(\bm\theta;\lambda)$ using starting point $i$ \Comment{Main text, Section~\ref{sec:param_infer}}
        \State Add $\est$ to $\estset$
    \EndFor
\EndFor
\State $\CRNset_\text{base}(\estset) \gets$ empty set of CRNs
\For{$\est\in\estset$}
    \State Determine set of CRNs $\CRNset_\text{prune}(\est)$ by pruning reactions with negligible dynamics \Comment{Section S2.1}
    \State $\CRNset_\text{base}(\estset) \gets \CRNset_\text{base}(\estset) \cup \CRNset_\text{prune}(\est)$ 
\EndFor
\State Generate exchange pairs $\pairset_\text{all}(\estset)$ from $\CRNset_\text{base}(\estset)$ \Comment{Equation \eqref{eq:all_pairs}}
\State Determine final CRN ensemble $\CRNset(\estset)$ by applying recombination to $\CRNset_\text{base}(\estset)$ based on $\pairset_\text{all}(\estset)$ \Comment{Section S2.2}
\State Compute $\textsc{bic}(\CRN)$ for each $\CRN\in\CRNset(\estset)$ \Comment{Section S2.3}
\State Approximate the posterior probability $p(\CRN\vert\data)$ for each $\CRN\in\CRNset(\estset)$ using (10) \Comment{Main text, Section~\ref{sec:posterior}}
\State \textbf{return} $\{p(\CRN\vert\data)\}_{\CRN\in\CRNset(\estset)}$
\end{algorithmic}
\end{algorithm}

\subsection{Pruning stage}

In Section~\ref{sec:map_ests} of the main text, we described that a parameter estimate $(\hat{\mathbf{k}},\hat{\bm\sigma}^2)=\est\in\estset$ can be mapped to a CRN by identifying the reactions $r$ with the largest values of 
\begin{equation}\label{eq:reaction_flux}
    g(r;\hat{\mathbf{k}}) = \int_0^{t_N} \hat{k}_r \prod_{s=1}^S \tilde{x}_s(t;\hat{\mathbf{k}})^{m_{rs}^-} \, dt,
\end{equation}
which quantifies the contribution of reaction $r$ to the system dynamics. Let $\CRN_\text{prune}(\est;m)$ denote the set of $m$ reactions corresponding to the $m$ highest values of $(g(r;\hat{\mathbf{k}}))_{r\in\fullCRN}$. Furthermore, let $\hat{\mathbf{k}}^{(m)}$ denote the vector which shares $m$ entries with $\hat{\mathbf{k}}$ corresponding to the reactions $\CRN_\text{prune}(\est;m)$, the other $\lvert\fullCRN\rvert-m$ entries of $\hat{\mathbf{k}}^{(m)}$ are zero. 

One way to map each parameter estimate $\est$ to a single CRN is to find the value of $m$ that results in the most parsimonious model fit. Formally, we can map $\est$ to $\CRN_\text{prune}(\est;m^*(\est))$, where
\begin{align*}
    m^*(\est) &\coloneqq \argmin_{m\in\{1,\ldots,\lvert\fullCRN\rvert\}} b(\est;m), \\
    b(\est;m) &\coloneqq -2 \log p(\data\vert\hat{\mathbf{k}}^{(m)},\hat{\bm\sigma}^2) + m \log\lvert\data\rvert .
\end{align*}
Note that $b(\est;m)$ resembles the BIC for the CRN $\CRN_\text{prune}(\est;m)$, though they are not exactly equal. This is because $\hat{\mathbf{k}}^{(m)}$ should be close, but not necessarily equal, to the maximum likelihood estimate for the CRN $\CRN_\text{prune}(\est;m)$. 

However, since we seek an ensemble of CRNs that plausibly explain the data, we do not need to restrict ourselves to mapping each estimate $\est$ to a single CRN only. Instead, let 
\begin{equation}\label{eq:multiple_m}
    \CRNset_\text{prune}(\est) \coloneqq \left\{ \CRN_\text{prune}(\est;m) \colon m=1,\ldots,\lvert\fullCRN\rvert \: \text{ such that } b(\est;m) < b(\est;m^*(\est)) + \delta \right\}
\end{equation}
for some threshold $\delta>0$; in this work we use $\delta=\log(10^6)$. $\CRNset_\text{prune}(\est)$ is a set of CRNs derived from the estimate $\est$ that we deem to potentially explain the data well without involving too many reactions. The base set of CRNs mentioned in Section~\ref{sec:map_ests} of the main text is defined as
\begin{equation*}
    \CRNset_\text{base}(\estset) \coloneqq \bigcup_{\est\in\estset} \CRNset_\text{prune}(\est).
\end{equation*}
A threshold $\delta$ in \eqref{eq:multiple_m} that is too low leads to false negatives, i.e. falsely excluding CRNs of appreciable posterior probability. Meanwhile, if $\delta$ is too high, $\CRNset_\text{base}(\estset)$ includes many CRNs of low posterior probability, which increases computational burden for downstream analyses. Our choice of $\delta=\log(10^6)$ is considered conservative; increasing $\delta$ further (more conservative) would only introduce CRNs of low posterior probability that are most likely outside of the 95\% HPD set.

We note that for each CRN $\CRN\in\CRNset_\text{base}(\estset)$, there may be multiple parameter estimates $\est\in\estset$ where $\CRN_\text{prune}(\est; \lvert\CRN\rvert) = \CRN$. In this case, we keep track of the estimate $\est$ that gives the smallest value of $b(\est;\lvert\CRN\rvert)$. Formally, let $\bm{\theta}_\text{map}(\CRN;\estset)$ be
\begin{equation*}
     \argmin_{\est\in\estset} b(\est;\lvert\CRN\rvert) \quad \text{such that } \CRN_\text{prune}(\est; \lvert\CRN\rvert) = \CRN
\end{equation*}
with rate constants corresponding to the reactions $\fullCRN\setminus\CRN$ set to zero, and let $\bm{\theta}_\text{map}(\CRN;\estset) \eqqcolon (\mathbf{k}_\text{map}(\CRN;\estset), \bm{\sigma}^2_\text{map}(\CRN;\estset))$. We will use ${\bm\theta}_\text{map}(\CRN;\estset)$ to propose CRNs in the recombination stage, and to initialise optimisation problems performed for finding maximum likelihood estimates.

\subsection{Recombination stage}

The recombination stage aims to find all CRNs that have significant posterior mass, but are not identified in $\CRNset_\text{base}(\estset)$. The first step of the recombination stage is to identify pairs of CRNs $\CRN^1,\CRN^2\in\CRNset_\text{base}(\estset)$ such that replacing the reactions ${\CRN^{2\setminus 1}\coloneqq\CRN^2\setminus \CRN^1}$ with the reactions ${\CRN^{1\setminus 2}\coloneqq\CRN^1\setminus \CRN^2}$ potentially results in a similar or better model fit. This motivates us to propose the CRN ${(\CRN\setminus\CRN^{2\setminus 1})\cup\CRN^{1\setminus 2}}$ for each CRN $\CRN\in\CRNset_\text{base}(\estset)$ that contains $\CRN^{2\setminus 1}$ as a subset. As an example, suppose there are $\lvert\fullCRN\rvert=5$ candidate reactions. Consider the CRNs
\begin{equation*}
    \CRN^1=\{2,3,5\}, \quad \CRN^2=\{4,5\}, \quad R=\{1,2,4\}.
\end{equation*}
This leads to $\CRN^{1\setminus 2} = \{2,3\}$ and $\CRN^{2\setminus 1} = \{4\}$, resulting in the proposal of the CRN ${(\CRN\setminus\CRN^{2\setminus 1})\cup\CRN^{1\setminus 2}}=\{1,2,3\}$.
We call the pair of reaction sets $(\CRN^{1\setminus 2},\CRN^{2\setminus 1})$ the \emph{exchange pair} induced by the pair of CRNs $(\CRN^1,\CRN^2)$. Let $\pairset_\text{all}(\estset)$ denote the set of exchange pairs generated by all pairs of CRNs from the pruning stage, formally defined as
\begin{equation}\label{eq:all_pairs}
    \pairset_\text{all}(\estset) \coloneqq \{ (\CRN^{1\setminus 2},\CRN^{2\setminus 1}) \colon \CRN^1,\CRN^2\in\CRNset_\text{base}(\estset) \: \text{ such that } \lvert\CRN^{1\setminus 2}\rvert \le 2, \lvert\CRN^{2\setminus 1}\rvert\le 2 \}.
\end{equation}
The size constraints on $\CRN^{1\setminus 2}$ and $\CRN^{2\setminus 1}$ are imposed for computational efficiency. Note that for each exchange pair $(U,V)\in\pairset_\text{all}(\estset)$, there may be multiple pairs of CRNs $R^1,R^2\in\CRNset_\text{base}(\estset)$ such that $(U,V) = (\CRN^{1\setminus 2},\CRN^{2\setminus 1})$. 

Generating recombined CRNs using all exchange pairs $\pairset_\text{all}(\estset)$ produces the set of CRNs
\begin{equation*}
    \CRNset_\text{all}(\estset) \coloneqq \{ (\CRN\setminus V)\cup U \colon \CRN\in\CRNset_\text{base}(\estset), \:(U,V)\in\pairset_\text{all}(\estset) \: \text{ such that } V \subseteq \CRN \}.
\end{equation*}
However, $\CRNset_\text{all}(\estset)$ consists of numerous CRNs, so calculating the BIC for each CRN in $\CRNset_\text{all}(\estset)$ would be computationally prohibitive, as each BIC calculation involves an optimisation problem. For the pyridine denitrogenation case study, we found that $\lvert \CRNset_\text{all}(\estset) \rvert > 2.6\times10^5$ for all penalty functions. Thus, we seek to efficiently identify promising CRN candidates from $\CRNset_\text{all}(\estset)$ without numerical optimisation. 

To do so, we need nominal values of $\bm{\theta}$ for each CRN $\CRN'\in\CRNset_\text{all}(\estset)$. Given some $\CRN'\in\CRNset_\text{all}(\estset)$, we find a nominal value of $\bm{\theta}$ for each way that $\CRN'$ can be expressed as ${(\CRN\setminus\CRN^{2\setminus 1})\cup\CRN^{1\setminus 2}}$ for some ${\CRN,\CRN^1,\CRN^2\in\CRNset_\text{base}(\estset)}$ satisfying $\lvert\CRN^{1\setminus 2}\rvert \le 2$, $\lvert\CRN^{2\setminus 1}\rvert\le 2$, and $\CRN^{2\setminus 1}\subseteq\CRN$. Let $\mathbf{k}_\text{nom}(\CRN,\CRN^1,\CRN^2;\estset)$ be the rate constants $\mathbf{k}_\text{map}(\CRN;\estset)$ with entries corresponding to the reactions $\CRN^{2\setminus 1}$ zeroed out, replaced with entries of $\mathbf{k}_\text{map}(\CRN^1;\estset)$ corresponding to the reactions $\CRN^{1\setminus 2}$. Formally, for each reaction $r\in\fullCRN$, the corresponding entry of $\mathbf{k}_\text{nom}(\CRN,\CRN^1,\CRN^2;\estset)$ is
\begin{equation*}
    \begin{cases}
        0 & \text{if } r\in\CRN^{2\setminus 1}, \\
        \text{entry $r$ of } \mathbf{k}_\text{map}(\CRN^1;\estset) & \text{if } r\in\CRN^{1\setminus 2}, \\
        \text{entry $r$ of } \mathbf{k}_\text{map}(\CRN;\estset) & \text{otherwise.}\\
    \end{cases}
\end{equation*}

Let us return to the example with a library of 5 candidate reactions and consider
\begin{equation*}
    \CRN^1=\{2,3,5\}, \quad \CRN^2=\{4,5\}, \quad R=\{1,2,4\}, \quad \CRN^{1\setminus 2} = \{2,3\}, \quad \CRN^{2\setminus 1} = \{4\}.
\end{equation*}
Suppose that $\mathbf{k}_\text{map}(\CRN;\estset)=(1,1,0,1,0)$ and $\mathbf{k}_\text{map}(\CRN^1;\estset)=(0,1.2,1.3,0,1.5)$. In this case, the nominal values of the rate constants of the proposed CRN ${(\CRN\setminus\CRN^{2\setminus 1})\cup\CRN^{1\setminus 2}}=\{1,2,3\}$ are ${\mathbf{k}_\text{nom}(\CRN,\CRN^1,\CRN^2;\estset)=(1,1.2,1.3,0,0)}$.

To determine nominal values of the noise variances, we find
\begin{equation}\label{eq:optim_variance}
    \bm{\sigma}^2_\text{nom}(\CRN,\CRN^1,\CRN^2;\estset) \coloneqq
    \argmax_{\bm{\sigma}^2} \log p(\data\vert \mathbf{k}_\text{nom}(\CRN,\CRN^1,\CRN^2;\estset), \bm{\sigma}^2 )
\end{equation}
while respecting the bounds described in Section~\ref{sec:opt_init_bound}. Since the likelihood is normal, the solution to \eqref{eq:optim_variance} can be found analytically (details omitted here). 

Given some CRN $R\in\CRNset_\text{base}(\estset)$ and exchange pair $(U,V)\in\pairset_\text{all}(\estset)$, we deem ${(\CRN\setminus V)\cup U}$ to be a promising alternative to $R$ if for some $\CRN^1,\CRN^2\in\CRNset_\text{base}(\estset)$ satisfying $(\CRN^{1\setminus 2}, \CRN^{2\setminus 1}) = (U, V)$, the log-likelihood of $({\mathbf{k}_\text{nom}(\CRN,\CRN^1,\CRN^2;\estset)},{\bm{\sigma}^2_\text{nom}(\CRN,\CRN^1,\CRN^2;\estset)})$ is similar to or greater than that of $\bm\theta_\text{map}(\CRN;\estset)$. For conciseness, we denote these two log-likelihoods as $l_\text{map}(\CRN;\estset)$ and $l_\text{nom}(\CRN,\CRN^1,\CRN^2;\estset)$ respectively. With this in mind, we seek to narrow down the exchange pairs $(U,V)\in\pairset_\text{all}(\estset)$ to those that are most likely to give similar or better model fits upon replacing reactions $V$ with reactions $U$. We rank all exchange pairs $(U,V)\in\pairset_\text{all}(\estset)$ in increasing order of
\begin{equation}\label{eq:rank_pairs}
\min_{\CRN^1,\CRN^2\in\CRNset_\text{base}(\estset)} \left[ l_\text{map}(\CRN^2;\estset) - l_\text{nom}(\CRN^2,\CRN^1,\CRN^2;\estset) \right] \quad \text{such that } (\CRN^{1\setminus 2},\CRN^{2\setminus 1}) = (U,V) .
\end{equation}
The rationale behind \eqref{eq:rank_pairs} is that if replacing reactions $\CRN^{2\setminus 1}$ in $\CRN^2$ with reactions $\CRN^{1\setminus 2}$ does not degrade the log-likelihood much, then $(\CRN^{1\setminus 2},\CRN^{2\setminus 1})$ is considered to be a promising exchange pair. Let $\pairset_\text{top}(\estset)$ denote the top 1000 exchange pairs ranked according to \eqref{eq:rank_pairs}. Generating recombined CRNs using the exchange pairs in $\pairset_\text{top}(\estset)$ produces the set of CRNs
\begin{equation*}
    \CRNset_\text{top}(\estset) \coloneqq \{ (\CRN\setminus V)\cup U \colon \CRN\in\CRNset_\text{base}(\estset), (U,V)\in\pairset_\text{top}(\estset) \text{ such that } V \subseteq \CRN \}.
\end{equation*}
We rank the CRNs $R' \in \CRNset_\text{top}(\estset)$ in decreasing order of
\begin{align}
    \max_{\CRN,\CRN^1,\CRN^2\in\CRNset_\text{base}(\estset)} &\left[ \log p(\CRN') + l_\text{nom}(\CRN,\CRN^1,\CRN^2;\estset) - \frac{1}{2}\lvert\CRN'\rvert \log \lvert\data\rvert \right] \label{eq:rank_crns} \\
    &\text{such that } \CRN' = (\CRN\setminus\CRN^{2\setminus 1})\cup\CRN^{1\setminus 2}, \: \CRN^{2\setminus 1} \subseteq \CRN, \: (\CRN^{1\setminus 2},\CRN^{2\setminus 1})\in\pairset_\text{top}(\estset). \nonumber
\end{align}
Note that the expression to be maximised in \eqref{eq:rank_crns} is analogous to the log unnormalised posterior (see numerator of \eqref{eq:bayes_crn_sel} in the main text), but $l_\text{nom}(\CRN,\CRN^1,\CRN^2;\estset)$ is the log-likelihood evaluated for some nominal value of $\bm\theta$, not the maximum log-likelihood. The CRN ensemble $\CRNset(\estset)$ mentioned in Section~\ref{sec:map_ests} of the main text is defined to be the top 1000 CRNs in $\CRNset_\text{top}(\estset)$ ranked according to \eqref{eq:rank_crns}.

\subsection{Finding maximum likelihood estimates}\label{sec:mle}

In the recombination stage, we have not optimised parameter values for the proposed CRNs. Our approximation of the model evidence is based on the BIC, which requires finding the maximum likelihood estimate; see \eqref{eq:bic} of the main text. For each $\CRN'\in\CRNset(\estset)$, we need to solve the optimisation problem
\begin{equation}
\begin{split}
    \textsc{bic}(\CRN') &= -2 \hat{L} + \lvert \CRN' \rvert \log \lvert \data \rvert,  \\
     \text{where } \hat{L} &= \max_{\bm\theta} \log p(\data\vert\mathbf{k},\bm\sigma^2) \quad \text{subject to } k_r = 0 \:\forall\, r \in \fullCRN\setminus\CRN'. 
\end{split}
\end{equation}
The optimisation details are akin to those described in Section~\ref{sec:optim}, with the exception of the starting points. Instead of using multiple starting points, we set the starting point of ${(\mathbf{k},\bm\sigma^2)}$ for CRN $R'\in\CRNset(\estset)$ to \splitatcommas{$(\mathbf{k}_\text{nom}(\CRN,\CRN^1,\CRN^2;\estset),\bm{\sigma}^2_\text{nom}(\CRN,\CRN^1,\CRN^2;\estset))$}, where ${(\CRN,\CRN^1,\CRN^2)}$ is the maximiser of \eqref{eq:rank_crns} given $\CRN'$.

\subsection{Construction of hierarchy trees}\label{sec:trees}

In this section, we present the formal construction of the hierarchy trees representing 95\% HPD sets $\CRNset^\text{HPD}$, which are depicted in Figures~3C, S5, and S7. Each node $v$ of a hierarchy tree is characterised by a set of included reactions $\CRN^\text{inc}(v)$ and a set of excluded reactions $\CRN^\text{exc}(v)$. Let $\CRNset(v)$ be the set of CRNs in $\CRNset^\text{HPD}$ that include reactions $\CRN^\text{inc}(v)$ and exclude reactions $\CRN^\text{exc}(v)$. Formally, we have
\begin{equation*}
    \CRNset(v) \coloneqq \{\CRN \in \CRNset^\text{HPD} \colon \CRN^\text{inc}(v) \subseteq \CRN, \: \CRN^\text{exc}(v) \cap \CRN = \varnothing \}.
\end{equation*}
The root node $v^\text{root}$ is associated with $\CRN^\text{inc}(v^\text{root})=\CRN^\text{exc}(v^\text{root})=\varnothing$, and thus $\CRNset(v^\text{root}) = \CRNset^\text{HPD}$. Each node $v$ is labelled by the posterior probability of $\CRNset(v)$, i.e. $\mathbb{P}(\CRN\in\CRNset(v)\vert\data)$, and the number of CRNs in $\CRNset(v)$ in brackets.

Each node $v$ has either 0, 1, or 2 child nodes. If node $v$ has 2 child nodes, its CRN set $\CRNset(v)$ is split into two and passed to its child nodes depending on the inclusion of the reaction $r^\text{mode}(v)$ (denoted as $r$ in the main text for brevity) with the highest posterior probability conditioned on the CRN set $\CRNset(v)$. The formal definition is
\begin{equation*}
    r^\text{mode}(v) = \argmax_{r\in \fullCRN\setminus\CRN^\text{inc}(v)} \mathbb{P}(r\in \CRN \vert \data, \CRN\in\CRNset(v));
\end{equation*}
ties are broken arbitrarily. This leads to recursive definitions of $\CRN^\text{inc}(\cdot)$ and $\CRN^\text{exc}(\cdot)$:
\begin{align*}
    \CRN^\text{inc}(\text{left child of }v) &= \CRN^\text{inc}(v) \cup \{r^\text{mode}(v)\}, & \CRN^\text{exc}(\text{left child of }v) &= \CRN^\text{exc}(v), \\
    \CRN^\text{inc}(\text{right child of }v) &= \CRN^\text{inc}(v), & \CRN^\text{exc}(\text{right child of }v) &= \CRN^\text{exc}(v) \cup \{r^\text{mode}(v)\}.
\end{align*}
If every CRN in $\CRNset(v)$ includes reaction $r^\text{mode}(v)$, then node $v$ has no right child, as its corresponding CRN set would be empty. Each edge emanating from node $v$ is labelled with the inclusion or exclusion of reaction $r^\text{mode}(v)$.

If there is only one CRN in $\CRNset(v)$ that includes reaction $r^\text{mode}(v)$, then node $v$ is considered a leaf node and has no child nodes. The leaf node is additionally labelled with the CRNs in $\CRNset(v)$, with reactions in $\CRN^\text{inc}(v)$ omitted. Note that this termination criterion always applies if $\CRNset(v)$ consists of a single CRN. The converse is not true: the termination criterion can also apply if $\CRNset(v)$ consists of multiple CRNs; see the leaf node $v$ labelled ``0.008(4)'' in Figure~3C, where \splitatcommas{$\CRNset(v) = \{\{13,16,19,1,2\}, \{13,16,19,1,4\}, \{13,16,19,1,5\}, \{13,16,19,1,24\}\}$}.

If $\CRN^\text{inc}(v) \in \CRNset(v)$, then node $v$ is also considered a leaf node and has no child nodes. The rationale here is that the CRNs in $\CRNset(v)$ other than $\CRN^\text{inc}(v)$ all feature the reactions in $\CRN^\text{inc}(v)$ as a strict subset, and thus are of less interest. These leaf nodes are differentiated from the leaf nodes following the previous termination criteria by the absence of additional node labels.

\clearpage

\section{Libraries of candidate reactions}

The generation of candidate reactions, their \LaTeX{} representations, and the corresponding ODE systems are implemented using the Julia package \texttt{Catalyst.jl}~\cite{loman_catalyst_2023}.

\subsection{Simulation study: synthetic CRN}

The library of candidate reactions $\fullCRN$ consists of
\begin{align*}
\tikzmark{sim1begin} X_1 &\xrightarrow{k_{1}} X_2 \tikzmark{sim1end} & X_3 &\xrightarrow{k_{11}} X_1 & X_2 + X_3 &\xrightarrow{k_{21}} X_1\\
X_1 &\xrightarrow{k_{2}} X_3 & X_3 &\xrightarrow{k_{12}} X_2 & X_2 + X_3 &\xrightarrow{k_{22}} X_2\\
X_1 &\xrightarrow{k_{3}} X_1 + X_2 & \tikzmark{sim13begin} X_3 &\xrightarrow{k_{13}} X_1 + X_2 \tikzmark{sim13end} & X_2 + X_3 &\xrightarrow{k_{23}} X_3\\
X_1 &\xrightarrow{k_{4}} X_2 + X_3 & X_3 &\xrightarrow{k_{14}} X_2 + X_3 & X_2 + X_3 &\xrightarrow{k_{24}} X_1 + X_2\\
X_1 &\xrightarrow{k_{5}} X_1 + X_3 & X_3 &\xrightarrow{k_{15}} X_1 + X_3 & X_2 + X_3 &\xrightarrow{k_{25}} X_1 + X_3\\
X_2 &\xrightarrow{k_{6}} X_1 & X_1 + X_2 &\xrightarrow{k_{16}} X_1 & X_1 + X_3 &\xrightarrow{k_{26}} X_1\\
X_2 &\xrightarrow{k_{7}} X_3 & X_1 + X_2 &\xrightarrow{k_{17}} X_2 & X_1 + X_3 &\xrightarrow{k_{27}} X_2\\
X_2 &\xrightarrow{k_{8}} X_1 + X_2 & \tikzmark{sim18begin} X_1 + X_2 &\xrightarrow{k_{18}} X_3 \tikzmark{sim18end} & X_1 + X_3 &\xrightarrow{k_{28}} X_3\\
X_2 &\xrightarrow{k_{9}} X_2 + X_3 & X_1 + X_2 &\xrightarrow{k_{19}} X_2 + X_3 & X_1 + X_3 &\xrightarrow{k_{29}} X_1 + X_2\\
X_2 &\xrightarrow{k_{10}} X_1 + X_3 & X_1 + X_2 &\xrightarrow{k_{20}} X_1 + X_3 & X_1 + X_3 &\xrightarrow{k_{30}} X_2 + X_3.
\end{align*}
The boxed reactions are the reactions present in the ground-truth CRN.
\begin{tikzpicture}[overlay, remember picture]
\draw[black,thick]([shift={(-0.3em,2.8ex)}]pic cs:sim1begin)rectangle([shift={(0.3em,-0.8ex)}]pic cs:sim1end);
\draw[black,thick]([shift={(-0.3em,2.8ex)}]pic cs:sim13begin)rectangle([shift={(0.3em,-0.8ex)}]pic cs:sim13end);
\draw[black,thick]([shift={(-0.3em,2.8ex)}]pic cs:sim18begin)rectangle([shift={(0.3em,-0.8ex)}]pic cs:sim18end);
\end{tikzpicture}

\subsection{Case study 1: \texorpdfstring{\textalpha}{α}-pinene isomerisation}

The library of candidate reactions $\fullCRN$ consists of
\begin{align*}
    X_1 &\xrightarrow{k_{1}} X_2 & X_3 &\xrightarrow{k_{7}} X_1 & X_5 &\xrightarrow{k_{13}} 2X_1\\
X_1 &\xrightarrow{k_{2}} X_3 & X_3 &\xrightarrow{k_{8}} X_2 & X_5 &\xrightarrow{k_{14}} 2X_2\\
X_1 &\xrightarrow{k_{3}} X_4 & X_3 &\xrightarrow{k_{9}} X_4 & X_5 &\xrightarrow{k_{15}} 2X_3\\
X_2 &\xrightarrow{k_{4}} X_1 & X_4 &\xrightarrow{k_{10}} X_1 & X_5 &\xrightarrow{k_{16}} 2X_4\\
X_2 &\xrightarrow{k_{5}} X_3 & X_4 &\xrightarrow{k_{11}} X_2 & 2X_1 &\xrightarrow{k_{17}} X_5\\
X_2 &\xrightarrow{k_{6}} X_4 & X_4 &\xrightarrow{k_{12}} X_3 & 2X_2 &\xrightarrow{k_{18}} X_5\\
 &  &  &  & 2X_3 &\xrightarrow{k_{19}} X_5\\
 &  &  &  & 2X_4 &\xrightarrow{k_{20}} X_5.
\end{align*}

\subsection{Case study 2: pyridine denitrogenation}\label{sec:pyridine_lib}

\begin{table}[ht]
\centering
\begin{tabular}{*{3}{c}}
\toprule
Species & Name & Molecular formula \\
\midrule
$X_1$ & Pyridine & $\mathrm{C_{5}H_{5}N}$ \\
$X_2$ & Piperidine & $\mathrm{C_{5}H_{11}N}$ \\
$X_3$ & Pentylamine & $\mathrm{C_{5}H_{13}N}$ \\
$X_4$ & N-pentylpiperidine & $\mathrm{C_{10}H_{21}N}$ \\
$X_5$ & Dipentylamine & $\mathrm{C_{10}H_{23}N}$ \\
$X_6$ & Ammonia & $\mathrm{NH_3}$ \\
$X_7$ & Pentane & $\mathrm{C_{5}H_{12}}$ \\
\bottomrule
\end{tabular}
\tblspace
\caption{List of species involved in pyridine denitrogenation~\cite{bock_numerical_1981}, with hydrogen ($\mathrm{H_2}$) excluded.}\label{tbl:pyridine_species}
\end{table}

The reaction scheme proposed in \cite{bock_numerical_1981} features the species in Table~\ref{tbl:pyridine_species} and hydrogen ($\mathrm{H_2}$). Hydrogen is not featured in the ODE system, which is appropriate if the system takes place in a hydrogen-rich environment. Pentane is not reported in the data~\cite{schittkowski_collection_2009}, though this is not an issue as pentane is featured only as a product in the reaction scheme. With this in mind, our library of candidate reactions is generated with the following assumptions:
\begin{enumerate}[label=(\roman*)]
    \item excluding $\mathrm{H_2}$, each reaction features 1 or 2 reactant molecules and 1 or 2 product molecules,
    \item no species is featured simultaneously as a reactant and a product in a reaction,
    \item pentane is not a reactant, and
    \item the numbers of carbon and nitrogen atoms are balanced for each reaction.
\end{enumerate}
Following these assumptions, we obtain a library of 67 candidate reactions. Since pentane ($X_7$) is not a reactant and thus does not affect the system dynamics, we omit it when displaying the candidate reactions $\fullCRN$:
\begin{align*}
\tikzmark{start1} X_1 &\xrightarrow{k_{1}} X_2 \tikzmark{end1} & 2X_1 &\xrightarrow{k_{18}} 2X_2 & X_1 + X_2 &\xrightarrow{k_{35}} 2X_3 & X_3 + X_4 &\xrightarrow{k_{52}} X_1 + X_5\\
X_1 &\xrightarrow{k_{2}} X_3 & 2X_1 &\xrightarrow{k_{19}} 2X_3 & X_1 + X_2 &\xrightarrow{k_{36}} X_4 + X_6 & X_3 + X_4 &\xrightarrow{k_{53}} X_2 + X_5\\
X_1 &\xrightarrow{k_{3}} X_6 & 2X_1 &\xrightarrow{k_{20}} X_2 + X_3 & X_1 + X_2 &\xrightarrow{k_{37}} X_5 + X_6 & X_3 + X_5 &\xrightarrow{k_{54}} X_1 + X_4\\
\tikzmark{start4} X_2 &\xrightarrow{k_{4}} X_1 \tikzmark{end4} & 2X_1 &\xrightarrow{k_{21}} X_4 + X_6 & X_1 + X_3 &\xrightarrow{k_{38}} 2X_2 & X_3 + X_5 &\xrightarrow{k_{55}} X_2 + X_4\\
\tikzmark{start5} X_2 &\xrightarrow{k_{5}} X_3 \tikzmark{end5} & 2X_1 &\xrightarrow{k_{22}} X_5 + X_6 & X_1 + X_3 &\xrightarrow{k_{39}} X_4 + X_6 & X_4 + X_6 &\xrightarrow{k_{56}} 2X_1\\
X_2 &\xrightarrow{k_{6}} X_6 & 2X_2 &\xrightarrow{k_{23}} 2X_1 & X_1 + X_3 &\xrightarrow{k_{40}} X_5 + X_6 & X_4 + X_6 &\xrightarrow{k_{57}} 2X_2\\
X_3 &\xrightarrow{k_{7}} X_1 & 2X_2 &\xrightarrow{k_{24}} 2X_3 & X_1 + X_4 &\xrightarrow{k_{41}} X_2 + X_5 & X_4 + X_6 &\xrightarrow{k_{58}} 2X_3\\
X_3 &\xrightarrow{k_{8}} X_2 & 2X_2 &\xrightarrow{k_{25}} X_1 + X_3 & X_1 + X_4 &\xrightarrow{k_{42}} X_3 + X_5 & X_4 + X_6 &\xrightarrow{k_{59}} X_1 + X_2\\
\tikzmark{start9} X_3 &\xrightarrow{k_{9}} X_6 \tikzmark{end9} & 2X_2 &\xrightarrow{k_{26}} X_4 + X_6 & X_1 + X_5 &\xrightarrow{k_{43}} X_2 + X_4 & X_4 + X_6 &\xrightarrow{k_{60}} X_1 + X_3\\
\tikzmark{start10} X_4 &\xrightarrow{k_{10}} X_5 \tikzmark{end10} & 2X_2 &\xrightarrow{k_{27}} X_5 + X_6 & X_1 + X_5 &\xrightarrow{k_{44}} X_3 + X_4 & \tikzmark{start61} X_4 + X_6 &\xrightarrow{k_{61}} X_2 + X_3 \tikzmark{end61}\\
X_4 &\xrightarrow{k_{11}} X_1 & 2X_3 &\xrightarrow{k_{28}} 2X_1 & X_2 + X_3 &\xrightarrow{k_{45}} 2X_1 & X_5 + X_6 &\xrightarrow{k_{62}} 2X_1\\
\tikzmark{start12} X_4 &\xrightarrow{k_{12}} X_2 \tikzmark{end12} & 2X_3 &\xrightarrow{k_{29}} 2X_2 & \tikzmark{start46} X_2 + X_3 &\xrightarrow{k_{46}} X_4 + X_6 \tikzmark{end46} & X_5 + X_6 &\xrightarrow{k_{63}} 2X_2\\
X_4 &\xrightarrow{k_{13}} X_3 & 2X_3 &\xrightarrow{k_{30}} X_1 + X_2 & X_2 + X_3 &\xrightarrow{k_{47}} X_5 + X_6 & \tikzmark{start64} X_5 + X_6 &\xrightarrow{k_{64}} 2X_3 \tikzmark{end64}\\
X_5 &\xrightarrow{k_{14}} X_4 & 2X_3 &\xrightarrow{k_{31}} X_4 + X_6 & X_2 + X_4 &\xrightarrow{k_{48}} X_1 + X_5 & X_5 + X_6 &\xrightarrow{k_{65}} X_1 + X_2\\
X_5 &\xrightarrow{k_{15}} X_1 & \tikzmark{start32} 2X_3 &\xrightarrow{k_{32}} X_5 + X_6 \tikzmark{end32} & X_2 + X_4 &\xrightarrow{k_{49}} X_3 + X_5 & X_5 + X_6 &\xrightarrow{k_{66}} X_1 + X_3\\
X_5 &\xrightarrow{k_{16}} X_2 & 2X_4 &\xrightarrow{k_{33}} 2X_5 & X_2 + X_5 &\xrightarrow{k_{50}} X_1 + X_4 & X_5 + X_6 &\xrightarrow{k_{67}} X_2 + X_3.\\
\tikzmark{start17} X_5 &\xrightarrow{k_{17}} X_3 \tikzmark{end17} & 2X_5 &\xrightarrow{k_{34}} 2X_4 & X_2 + X_5 &\xrightarrow{k_{51}} X_3 + X_4 &  & 
\end{align*}
The boxed reactions are the reactions present in the gold-standard CRN proposed in \cite{bock_numerical_1981}.
\begin{tikzpicture}[overlay, remember picture]
\draw[black,thick]([shift={(-0.3em,2.8ex)}]pic cs:start1)rectangle([shift={(0.3em,-0.8ex)}]pic cs:end1);
\draw[black,thick]([shift={(-0.3em,2.8ex)}]pic cs:start5)rectangle([shift={(0.3em,-0.8ex)}]pic cs:end5);
\draw[black,thick]([shift={(-0.3em,2.8ex)}]pic cs:start46)rectangle([shift={(0.3em,-0.8ex)}]pic cs:end46);
\draw[black,thick]([shift={(-0.3em,2.8ex)}]pic cs:start32)rectangle([shift={(0.3em,-0.8ex)}]pic cs:end32);
\draw[black,thick]([shift={(-0.3em,2.8ex)}]pic cs:start10)rectangle([shift={(0.3em,-0.8ex)}]pic cs:end10);
\draw[black,thick]([shift={(-0.3em,2.8ex)}]pic cs:start9)rectangle([shift={(0.3em,-0.8ex)}]pic cs:end9);
\draw[black,thick]([shift={(-0.3em,2.8ex)}]pic cs:start12)rectangle([shift={(0.3em,-0.8ex)}]pic cs:end12);
\draw[black,thick]([shift={(-0.3em,2.8ex)}]pic cs:start17)rectangle([shift={(0.3em,-0.8ex)}]pic cs:end17);
\draw[black,thick]([shift={(-0.3em,2.8ex)}]pic cs:start4)rectangle([shift={(0.3em,-0.8ex)}]pic cs:end4);
\draw[black,thick]([shift={(-0.3em,2.8ex)}]pic cs:start61)rectangle([shift={(0.3em,-0.8ex)}]pic cs:end61);
\draw[black,thick]([shift={(-0.3em,2.8ex)}]pic cs:start64)rectangle([shift={(0.3em,-0.8ex)}]pic cs:end64);
\end{tikzpicture}

\clearpage

\section{Structural distinguishability}

\subsection{Decomposition result for structurally indistinguishable reaction sets}

Recall that a pair of structurally indistinguishable reaction sets $(\CRN,\CRN')$ is \emph{elementary} if 
\begin{enumerate}[label=(\roman*)]
    \item $\CRN$ and $\CRN'$ are not both empty, and 
    \item for any subsets $\CRN^*\subseteq \CRN$ and ${\CRN'}^*\subseteq \CRN'$ that are structurally indistinguishable and not both empty, we must have $(\CRN^*,{\CRN'}^*)=(\CRN, \CRN')$.
\end{enumerate}
Note that for any reaction~$r$ with a nonzero reaction vector, we have the \emph{trivial} elementary pair of structurally indistinguishable reaction singletons $(\{r\}, \{r\})$. All other elementary pairs of structurally indistinguishable reaction sets, which we call \emph{non-trivial} elementary pairs, must therefore be disjoint reaction sets as they cannot contain any trivial elementary pairs by condition (ii).

We will show in the proof of Theorem~\ref{thm:restate} that elementary pairs of structurally indistinguishable reaction sets correspond to \emph{elementary vectors}~\cite{rockafellarElementaryVectors1967} of a particular subspace. A vector $\mathbf{v}\in\mathbb{R}^D$ of a linear subspace $V\subseteq\mathbb{R}^D$ is \emph{elementary} if its support $\mathrm{supp}(\mathbf{v})$ (i.e. the indices of the nonzero entries of $\mathbf{v}$) is \emph{minimal}, that is, $\mathrm{supp}(\mathbf{v})$ does not properly contain the support of any nonzero vector in $V$. We say that vector $\mathbf{v} = (v_j)_{j=1}^D$ \emph{conforms to} vector $\mathbf{u} = (u_j)_{j=1}^D$ if for every $j=1,\ldots,D$, we have that
\begin{align*}
    u_j = 0 &\implies v_j = 0, \\
    u_j > 0 &\implies v_j \ge 0, \\
    u_j < 0 &\implies v_j \le 0.
\end{align*}
See \textcite{mullerElementaryVectors2016} for an accessible exposition on elementary vectors and conformal decompositions. To prove Theorem~\ref{thm:elementary} of the main text, we will make use of the following result:

\begin{theorem}[\textcite{rockafellarElementaryVectors1967}]\label{thm:conformal}
    Every vector~$\mathbf{u}\in V$ can be written as a sum of elementary vectors of $V$ that each conform to $\mathbf{u}$.
\end{theorem}
\begin{proof}
    See Theorem 1 of \textcite{rockafellarElementaryVectors1967} or Theorem 3 of \textcite{mullerElementaryVectors2016}.
\end{proof}

We give a proof of the following decomposition result on structurally indistinguishable reaction sets:

\renewcommand{\thetheorem}{\arabic{theorem}}
\begin{theorem}[restated from main text]\label{thm:restate}
Every pair of structurally indistinguishable reaction sets $(\CRN, \CRN')$ can be written as a componentwise union of elementary pairs of structurally indistinguishable reaction sets $(\CRN_1, \CRN'_1), \ldots, (\CRN_n, \CRN'_n)$:
    \begin{equation}\label{eq:supp_decomp}
        (\CRN, \CRN') = \biggl( \bigcup_{i=1}^n \CRN_i, \bigcup_{i=1}^n \CRN'_i \biggr).
    \end{equation}
\end{theorem}
\begin{proof}
    Since structural indistinguishability is a property that decomposes over reactant complexes (see Theorem~\ref{thm:distinguish} of the main text), it suffices to prove the result for the case where all reactions in $\CRN \cup \CRN'$ share the same reactant complex. Specifically, to construct the decomposition in \eqref{eq:supp_decomp} for the general case, we would partition the reactions in $\CRN$ and $\CRN'$ according to their reactant complexes, apply the single-complex result to each part of the partition to obtain a decomposition into elementary pairs supported on that complex, and combine the resulting decompositions across complexes. For the remainder of the proof, we assume that all reactions in $\CRN \cup \CRN'$ share the same reactant complex.
    
    By Theorem~\ref{thm:distinguish} of the main text, the structural indistinguishability of $\CRN$ and $\CRN'$ is equivalent to the existence of positive reals $(u_r)_{r\in\CRN}$ and $(u'_{r'})_{r'\in\CRN'}$ such that 
    \begin{equation}\label{eq:supp_SID}
        \sum_{r\in\CRN} u_r \mathbf{m}_r = \sum_{r'\in\CRN'} u'_{r'} \mathbf{m}_{r'},
    \end{equation}
    where $\mathbf{m}_r$ is the reaction vector of reaction $r$ (see Section~\ref{sec:prelims_CRNs} of the main text). Let $\mathbf{u}\in\mathbb{R}_+^{\lvert \CRN \rvert + \lvert \CRN' \rvert}$ be the vector concatenation of $(u_r)_{r\in\CRN}$ and $(u'_{r'})_{r'\in\CRN'}$, and let $\mathbf{A}$ be a $S\times (\lvert \CRN \rvert + \lvert \CRN' \rvert)$ matrix formed by horizontally concatenating $(\mathbf{m}_r)_{r\in\CRN}$ and $(-\mathbf{m}_{r'})_{r'\in\CRN'}$. The equality~\eqref{eq:supp_SID} is equivalent to stating that $\mathbf{u}$ is in the nullspace of $\mathbf{A}$, which we denote as $\mathrm{null}(\mathbf{A})$. For the same reason, each nonnegative vector $\mathbf{v}\in\mathrm{null}(\mathbf{A})$ corresponds to a pair of structurally indistinguishable reaction sets $(\CRN^*, {\CRN'}^*)$ satisfying $\CRN^*\subseteq \CRN$ and ${\CRN'}^*\subseteq \CRN'$. Specifically, the reactions in $\CRN^*$ correspond to the subset of the first $\lvert \CRN \rvert$ entries of $\mathbf{v}$ that are positive, while the reactions in ${\CRN'}^*$ correspond to the subset of the last $\lvert \CRN' \rvert$ entries of $\mathbf{v}$ that are positive. Moreover, the pair $(\CRN^*, {\CRN'}^*)$ is elementary if $\mathbf{v}$ is an elementary vector of $\mathrm{null}(\mathbf{A})$, since the support-minimal property of elementary vectors implies that any nonzero vector in $\mathrm{null}(\mathbf{A})$ whose support is contained in $\mathrm{supp}(\mathbf{v})$ must have support identical to $\mathrm{supp}(\mathbf{v})$.

    According to Theorem~\ref{thm:conformal}, there exist elementary vectors $\mathbf{v}_1,\ldots,\mathbf{v}_n$ of $\mathrm{null}(\mathbf{A})$ that each conform to $\mathbf{u}$ satisfying
    \begin{equation}\label{supp:vec_sum}
        \mathbf{u} = \mathbf{v}_1 + \cdots + \mathbf{v}_n.
    \end{equation}
    Since all entries of $\mathbf{u}$ are positive, it follows that all entries of $\mathbf{v}_i$ are nonnegative for each $i=1,\ldots,n$. Thus, $\mathbf{v}_i$ corresponds to an elementary pair of structurally indistinguishable reaction sets $(\CRN_i, \CRN'_i)$ where $\CRN_i \subseteq \CRN$ and $\CRN'_i \subseteq \CRN'$. For each reaction $r\in \CRN$, there must be some $\mathbf{v}_i$ whose corresponding entry is positive, as the sum of these entries is positive according to \eqref{supp:vec_sum}; the same holds for reactions in $\CRN'$. It follows that the pair $(\CRN, \CRN')$ is the componentwise union of the elementary pairs $(\CRN_1, \CRN'_1), \ldots, (\CRN_n, \CRN'_n)$.
\end{proof}

\subsection{Enumerating elementary pairs}

Having established that elementary pairs of structurally indistinguishable reaction sets are the building blocks of structurally indistinguishable CRNs, we now address the problem of enumerating these elementary pairs given a library~$\fullCRN$ of candidate reactions. Following the proof of Theorem~\ref{thm:restate}, we note that each elementary pair of structurally indistinguishable reaction sets involved in the decomposition~\eqref{eq:supp_decomp} consists of reactions that share the same reactant complex. It follows from condition~(ii) that the decomposition of an elementary pair must be the elementary pair itself; thus, every elementary pair consists of reactions that share the same reactant complex. 

\review{For each complex $\cplx$, let $\CRN^\cplx$ denote the set of reactions with $\cplx$ as the reactant complex, let $\mathbf{A}$ be the $S\times 2\lvert \CRN^\cplx \rvert$ matrix formed by horizontally concatenating $(\mathbf{m}_r)_{r\in\CRN^\cplx}$ and $(-\mathbf{m}_{r})_{r\in\CRN^\cplx}$, and let $\mathcal{P}$ denote the polyhedral cone $\textrm{null}(\mathbf{A})\cap\mathbb{R}^{2\lvert\CRN^\cplx\rvert}_{\ge 0}$. By \eqref{eq:supp_SID}, every nonzero $\mathbf{v}\in \mathcal{P}$ corresponds to a pair of structurally indistinguishable subsets of $\CRN^\cplx$, given by the positive entries among the first and last $\lvert\CRN^\cplx\rvert$ entries of $\mathbf{v}$. Conversely, every pair of structurally indistinguishable subsets of $\CRN^\cplx$ can be mapped to a nonzero vector of $\mathcal{P}$ by the same mechanism. It follows from condition~(ii) that the elementary pairs that involve~$\cplx$ correspond to the support-minimal vectors of $\mathcal{P}$. Because $\mathcal{P}$ lies in the nonnegative orthant, these vectors span exactly its extreme rays (Proposition~5 and Section~2.3 of \textcite{mullerElementaryVectors2016}), which can be enumerated with the \emph{double description} algorithm~\cite{motzkinDoubleDescription1953,fukudaDoubleDescription1996}. This algorithm has also been used in the CRN literature to find elementary flux modes~\cite{terzerLargescaleComputation2008}. Our notion of elementary pairs is distinct from that of elementary flux modes, although the mathematical objects involved are closely related.

The upshot is that we can enumerate the extreme rays of the polyhedral cone $\mathcal{P}$ via the double description algorithm~\cite{motzkinDoubleDescription1953}, which are subsequently mapped to elementary pairs by inspecting the support of the extreme rays obtained. Performing this for each reactant complex amounts to enumerating all elementary pairs. This procedure is outlined in Algorithm~\ref{alg:enumerate}.}

\begin{algorithm}
\caption{Enumeration of elementary pairs of structurally indistinguishable reaction sets}\label{alg:enumerate}
\begin{algorithmic}[1]
\Require{Library of candidate reactions $\fullCRN$, set of complexes $\cplxset$}
\For{$\cplx \in \cplxset$}
    \State $\mathbf{M} \gets (\mathbf{m}_r)_{r\in\CRN^\cplx}$
    \State $\mathbf{A} \gets [\mathbf{M} \enspace -\mathbf{M}]$
    \review{\State $\mathcal{E} \gets$ set of extremal rays of the polyhedral cone $\textrm{null}(\mathbf{A})\cap\mathbb{R}^{2\lvert\CRN^\cplx\rvert}_{\ge 0}$ obtained via the double description algorithm}
    \For{$\mathbf{v} \in \mathcal{E}$}        
        \State $\CRN \gets$ reactions corresponding to the subset of the first $\lvert \CRN^\cplx \rvert$ entries of $\mathbf{v}$ that are positive
        \State $\CRN' \gets$ reactions corresponding to the subset of the last $\lvert \CRN^\cplx \rvert$ entries of $\mathbf{v}$ that are positive
        \State \textbf{output} $(\CRN, \CRN')$ as an elementary pair
    \EndFor
\EndFor
\end{algorithmic}
\end{algorithm}

\subsection{Non-trivial elementary pairs for examples from main text}

We use Algorithm~\ref{alg:enumerate} to list the non-trivial elementary pairs of structurally indistinguishable reaction sets for each example presented in the main text. 

\subsubsection{Simulation study: synthetic CRN}

The boxed pairs correspond to elementary pairs where one of the reaction sets is a subset of the reactions in the ground-truth CRN.

Reactant complex $X_1$:
\begin{alignat*}{2}
    \bigl(\{X_1 \xrightarrow{k_{2}} X_3,\, X_1 \xrightarrow{k_{3}} X_1 + X_2&\}, \enspace &&\{X_1 \xrightarrow{k_{4}} X_2 + X_3\}\bigr) \\
    \bigl(\{X_1 \xrightarrow{k_{1}} X_2,\, X_1 \xrightarrow{k_{5}} X_1 + X_3&\}, \enspace &&\{X_1 \xrightarrow{k_{4}} X_2 + X_3\}\bigr) \\
    \bigl(\{X_1 \xrightarrow{k_{1}} X_2,\, X_1 \xrightarrow{k_{5}} X_1 + X_3&\}, \enspace &&\{X_1 \xrightarrow{k_{2}} X_3,\, X_1 \xrightarrow{k_{3}} X_1 + X_2\}\bigr)
\end{alignat*}

Reactant complex $X_2$:
\begin{alignat*}{2}
    \bigl(\{X_2 \xrightarrow{k_{7}} X_3,\, X_2 \xrightarrow{k_{8}} X_1 + X_2&\}, \enspace &&\{X_2 \xrightarrow{k_{10}} X_1 + X_3\}\bigr) \\
    \bigl(\{X_2 \xrightarrow{k_{6}} X_1,\, X_2 \xrightarrow{k_{9}} X_2 + X_3&\}, \enspace &&\{X_2 \xrightarrow{k_{10}} X_1 + X_3\}\bigr) \\
    \bigl(\{X_2 \xrightarrow{k_{6}} X_1,\, X_2 \xrightarrow{k_{9}} X_2 + X_3&\}, \enspace &&\{X_2 \xrightarrow{k_{7}} X_3,\, X_2 \xrightarrow{k_{8}} X_1 + X_2\}\bigr)
\end{alignat*}

Reactant complex $X_3$:
\begin{alignat*}{2}
    \tikzmark{epair1begin} \bigl(\{X_3 \xrightarrow{k_{11}} X_1,\, X_3 \xrightarrow{k_{14}} X_2 + X_3&\}, \enspace &&\{X_3 \xrightarrow{k_{13}} X_1 + X_2\}\bigr) \tikzmark{epair1end} \\
    \tikzmark{epair2begin} \bigl(\{X_3 \xrightarrow{k_{12}} X_2,\, X_3 \xrightarrow{k_{15}} X_1 + X_3&\}, \enspace &&\{X_3 \xrightarrow{k_{13}} X_1 + X_2\}\bigr) \tikzmark{epair2end} \\
    \bigl(\{X_3 \xrightarrow{k_{11}} X_1,\, X_3 \xrightarrow{k_{14}} X_2 + X_3&\}, \enspace &&\{X_3 \xrightarrow{k_{12}} X_2,\, X_3 \xrightarrow{k_{15}} X_1 + X_3\}\bigr)
\end{alignat*}
\begin{tikzpicture}[overlay, remember picture]
\draw[black,thick]([shift={(-0.3em,2.6ex)}]pic cs:epair1begin)rectangle([shift={(0.3em,-1.0ex)}]pic cs:epair1end);
\draw[black,thick]([shift={(-0.3em,2.6ex)}]pic cs:epair2begin)rectangle([shift={(0.3em,-1.0ex)}]pic cs:epair2end);
\end{tikzpicture}

Reactant complex $X_1+X_2$:
\begin{alignat*}{2}
    \tikzmark{epair3begin} \bigl(\{X_1+X_2 \xrightarrow{k_{16}} X_1,\, X_1+X_2 \xrightarrow{k_{19}} X_2 + X_3&\}, \enspace &&\{X_1+X_2 \xrightarrow{k_{18}} X_3\}\bigr) \tikzmark{epair3end} \\
    \tikzmark{epair4begin} \bigl(\{X_1+X_2 \xrightarrow{k_{17}} X_2,\, X_1+X_2 \xrightarrow{k_{20}} X_1 + X_3&\}, \enspace &&\{X_1+X_2 \xrightarrow{k_{18}} X_3\}\bigr) \tikzmark{epair4end} \\
    \bigl(\{X_1+X_2 \xrightarrow{k_{16}} X_1,\, X_1+X_2 \xrightarrow{k_{19}} X_2 + X_3&\}, \enspace &&\{X_1+X_2 \xrightarrow{k_{17}} X_2,\, X_1+X_2 \xrightarrow{k_{20}} X_1 + X_3\}\bigr)
\end{alignat*}
\begin{tikzpicture}[overlay, remember picture]
\draw[black,thick]([shift={(-0.3em,2.6ex)}]pic cs:epair3begin)rectangle([shift={(0.3em,-1.0ex)}]pic cs:epair3end);
\draw[black,thick]([shift={(-0.3em,2.6ex)}]pic cs:epair4begin)rectangle([shift={(0.3em,-1.0ex)}]pic cs:epair4end);
\end{tikzpicture}

Reactant complex $X_2+X_3$:
\begin{alignat*}{2}
    \bigl(\{X_2+X_3 \xrightarrow{k_{21}} X_1&\}, \enspace &&\{X_2+X_3 \xrightarrow{k_{22}} X_2,\, X_2+X_3 \xrightarrow{k_{25}} X_1 + X_3\}\bigr) \\
    \bigl(\{X_2+X_3 \xrightarrow{k_{21}} X_1&\}, \enspace &&\{X_2+X_3 \xrightarrow{k_{23}} X_3,\, X_2+X_3 \xrightarrow{k_{24}} X_1 + X_2\}\bigr) \\
    \bigl(\{X_2+X_3 \xrightarrow{k_{22}} X_2,\, X_2+X_3 \xrightarrow{k_{25}} X_1 + X_3&\}, \enspace &&\{X_2+X_3 \xrightarrow{k_{23}} X_3,\, X_2+X_3 \xrightarrow{k_{24}} X_1 + X_2\}\bigr)
\end{alignat*}

Reactant complex $X_1+X_3$:
\begin{alignat*}{2}
    \bigl(\{X_1+X_3 \xrightarrow{k_{26}} X_1,\, X_1+X_3 \xrightarrow{k_{30}} X_2 + X_3&\}, \enspace &&\{X_1+X_3 \xrightarrow{k_{27}} X_2\}\bigr) \\
    \bigl(\{X_1+X_3 \xrightarrow{k_{27}} X_2&\}, \enspace &&\{X_1+X_3 \xrightarrow{k_{28}} X_3,\, X_1+X_3 \xrightarrow{k_{29}} X_1 + X_2\}\bigr) \\
    \bigl(\{X_1+X_3 \xrightarrow{k_{26}} X_1,\, X_1+X_3 \xrightarrow{k_{30}} X_2 + X_3&\}, \enspace &&\{X_1+X_3 \xrightarrow{k_{28}} X_3,\, X_1+X_3 \xrightarrow{k_{29}} X_1 + X_2\}\bigr)
\end{alignat*}

\subsubsection{Case study 1: \texorpdfstring{\textalpha}{α}-pinene isomerisation}

Algorithm~\ref{alg:enumerate} returns only trivial elementary pairs of structurally indistinguishable reaction sets for this example. There are no non-trivial elementary pairs, since for each complex~$\cplx$, the reaction vectors corresponding to the reactions with reactant complex~$\cplx$, namely $\{\mathbf{m}_r \colon \cplx^-_r = \cplx\}$, are linearly independent.

\subsubsection{Case study 2: pyridine denitrogenation}

The boxed pair is the only elementary pair where one of the reaction sets is a subset of the reactions in the gold-standard CRN proposed in~\cite{bock_numerical_1981}. However, this pair is not of much interest as it simply splits a production of $X_2+X_3$ into $2X_2$ and $2X_3$ separately.

Reactant complex $2X_1$:
\begin{alignat*}{2}
    \bigl(\{2X_1 \xrightarrow{k_{18}} 2X_2, 2X_1 \xrightarrow{k_{19}} 2X_3&\}, \enspace &&\{2X_1 \xrightarrow{k_{20}} X_2 + X_3\}\bigr) \\
\end{alignat*}

Reactant complex $2X_2$:
\begin{alignat*}{2}
    \bigl(\{2X_2 \xrightarrow{k_{23}} 2X_1, 2X_2 \xrightarrow{k_{24}} 2X_3&\}, \enspace &&\{2X_2 \xrightarrow{k_{25}} X_1 + X_3\}\bigr) \\
\end{alignat*}

Reactant complex $2X_3$:
\begin{alignat*}{2}
    \bigl(\{2X_3 \xrightarrow{k_{28}} 2X_1, 2X_3 \xrightarrow{k_{29}} 2X_2&\}, \enspace &&\{2X_3 \xrightarrow{k_{30}} X_1 + X_2\}\bigr) \\
\end{alignat*}

Reactant complex $X_4+X_6$:
\begin{alignat*}{2}
    \bigl(\{X_4+X_6 \xrightarrow{k_{56}} 2X_1,\, X_4+X_6 \xrightarrow{k_{57}} 2X_2&\}, \enspace &&\{X_4+X_6 \xrightarrow{k_{59}} X_1 + X_2\}\bigr) \\
    \bigl(\{X_4+X_6 \xrightarrow{k_{56}} 2X_1,\, X_4+X_6 \xrightarrow{k_{58}} 2X_3&\}, \enspace &&\{X_4+X_6 \xrightarrow{k_{60}} X_1 + X_3\}\bigr) \\
    \tikzmark{epair5begin} \bigl(\{X_4+X_6 \xrightarrow{k_{57}} 2X_2,\, X_4+X_6 \xrightarrow{k_{58}} 2X_3&\}, \enspace &&\{X_4+X_6 \xrightarrow{k_{61}} X_2 + X_3\}\bigr) \tikzmark{epair5end} \\
    \bigl(\{X_4+X_6 \xrightarrow{k_{58}} 2X_3,\, X_4+X_6 \xrightarrow{k_{59}} X_1 + X_2&\}, \enspace &&\{X_4+X_6 \xrightarrow{k_{60}} X_1 + X_3,\, X_4+X_6 \xrightarrow{k_{61}} X_2 + X_3\}\bigr) \\
    \bigl(\{X_4+X_6 \xrightarrow{k_{57}} 2X_2,\, X_4+X_6 \xrightarrow{k_{60}} X_1 + X_3&\}, \enspace &&\{X_4+X_6 \xrightarrow{k_{58}} 2X_3,\, X_4+X_6 \xrightarrow{k_{59}} X_1 + X_2\}\bigr) \\
    \bigl(\{X_4+X_6 \xrightarrow{k_{56}} 2X_1,\, X_4+X_6 \xrightarrow{k_{61}} X_2 + X_3&\}, \enspace &&\{X_4+X_6 \xrightarrow{k_{58}} 2X_3,\, X_4+X_6 \xrightarrow{k_{59}} X_1 + X_2\}\bigr) \\
    \bigl(\{X_4+X_6 \xrightarrow{k_{56}} 2X_1,\, X_4+X_6 \xrightarrow{k_{61}} X_2 + X_3&\}, \enspace &&\{X_4+X_6 \xrightarrow{k_{57}} 2X_2,\, X_4+X_6 \xrightarrow{k_{60}} X_1 + X_3\}\bigr) \\
    \bigl(\{X_4+X_6 \xrightarrow{k_{56}} 2X_1,\, X_4+X_6 \xrightarrow{k_{61}} X_2 + X_3&\}, \enspace &&\{X_4+X_6 \xrightarrow{k_{59}} X_1 + X_2,\, X_4+X_6 \xrightarrow{k_{60}} X_1 + X_3\}\bigr) \\
    \bigl(\{X_4+X_6 \xrightarrow{k_{57}} 2X_2,\, X_4+X_6 \xrightarrow{k_{60}} X_1 + X_3&\}, \enspace &&\{X_4+X_6 \xrightarrow{k_{59}} X_1 + X_2,\, X_4+X_6 \xrightarrow{k_{61}} X_2 + X_3\}\bigr) \\
\end{alignat*}
\begin{tikzpicture}[overlay, remember picture]
\draw[black,thick]([shift={(-0.3em,2.6ex)}]pic cs:epair5begin)rectangle([shift={(0.3em,-1.0ex)}]pic cs:epair5end);
\end{tikzpicture}

Reactant complex $X_5+X_6$:
\begin{alignat*}{2}
    \bigl(\{X_5+X_6 \xrightarrow{k_{62}} 2X_1,\, X_5+X_6 \xrightarrow{k_{63}} 2X_2&\}, \enspace &&\{X_5+X_6 \xrightarrow{k_{65}} X_1 + X_2\}\bigr) \\
    \bigl(\{X_5+X_6 \xrightarrow{k_{62}} 2X_1,\, X_5+X_6 \xrightarrow{k_{64}} 2X_3&\}, \enspace &&\{X_5+X_6 \xrightarrow{k_{66}} X_1 + X_3\}\bigr) \\
    \bigl(\{X_5+X_6 \xrightarrow{k_{63}} 2X_2,\, X_5+X_6 \xrightarrow{k_{64}} 2X_3&\}, \enspace &&\{X_5+X_6 \xrightarrow{k_{67}} X_2 + X_3\}\bigr) \\
    \bigl(\{X_5+X_6 \xrightarrow{k_{64}} 2X_3,\, X_5+X_6 \xrightarrow{k_{65}} X_1 + X_2&\}, \enspace &&\{X_5+X_6 \xrightarrow{k_{66}} X_1 + X_3,\, X_5+X_6 \xrightarrow{k_{67}} X_2 + X_3\}\bigr) \\
    \bigl(\{X_5+X_6 \xrightarrow{k_{63}} 2X_2,\, X_5+X_6 \xrightarrow{k_{66}} X_1 + X_3&\}, \enspace &&\{X_5+X_6 \xrightarrow{k_{64}} 2X_3,\, X_5+X_6 \xrightarrow{k_{65}} X_1 + X_2\}\bigr) \\
    \bigl(\{X_5+X_6 \xrightarrow{k_{62}} 2X_1,\, X_5+X_6 \xrightarrow{k_{67}} X_2 + X_3&\}, \enspace &&\{X_5+X_6 \xrightarrow{k_{64}} 2X_3,\, X_5+X_6 \xrightarrow{k_{65}} X_1 + X_2\}\bigr) \\
    \bigl(\{X_5+X_6 \xrightarrow{k_{62}} 2X_1,\, X_5+X_6 \xrightarrow{k_{67}} X_2 + X_3&\}, \enspace &&\{X_5+X_6 \xrightarrow{k_{63}} 2X_2,\, X_5+X_6 \xrightarrow{k_{66}} X_1 + X_3\}\bigr) \\
    \bigl(\{X_5+X_6 \xrightarrow{k_{62}} 2X_1,\, X_5+X_6 \xrightarrow{k_{67}} X_2 + X_3&\}, \enspace &&\{X_5+X_6 \xrightarrow{k_{65}} X_1 + X_2,\, X_5+X_6 \xrightarrow{k_{66}} X_1 + X_3\}\bigr) \\
    \bigl(\{X_5+X_6 \xrightarrow{k_{63}} 2X_2,\, X_5+X_6 \xrightarrow{k_{66}} X_1 + X_3&\}, \enspace &&\{X_5+X_6 \xrightarrow{k_{65}} X_1 + X_2,\, X_5+X_6 \xrightarrow{k_{67}} X_2 + X_3\}\bigr) \\
\end{alignat*}

\clearpage

\section{Supplementary results}
\renewcommand{\topfraction}{.95}
\renewcommand{\textfraction}{.05}
\renewcommand{\floatpagefraction}{.95}

\subsection{Simulation study: synthetic CRN}

\begin{figure}[h!]
\centering
\includegraphics[width = \textwidth]{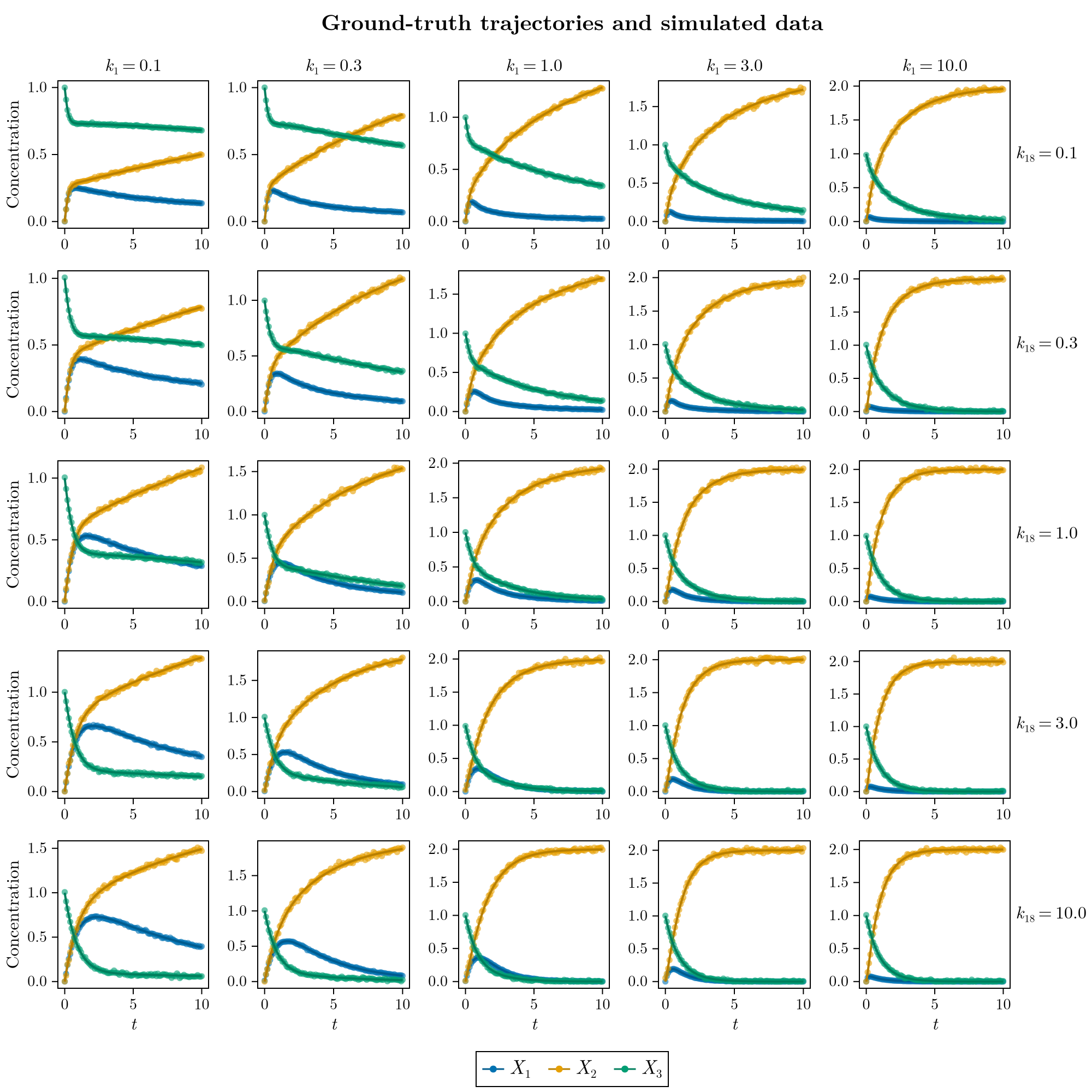}
\caption{Ground-truth trajectories and simulated datasets for the simulation study.}\label{fig:toy_data}
\end{figure}

We show in Figure~\ref{fig:toy_data} the 25 datasets used in our simulation study, each simulated with different ground-truth rate constants $\mathbf{k}^\text{true}$. Due to the use of additive Gaussian noise, simulated observations could be negative, which we truncate to zero. Recall that for each $\CRN\in\CRNset(\estset)$, we find the maximum likelihood estimate (MLE) of $\CRN$ (Section~\ref{sec:mle}). To quantify the trajectory reconstruction error of an MLE $\mathbf{k}^\text{MLE}$ (Figure~\ref{fig:toy_mode_traj} of the main text), we calculate the time-averaged absolute differences between the ground-truth and reconstructed trajectories, summed over species:
\begin{equation}\label{eq:err}
    \frac{1}{t_N}\int_0^{t_N} \sum_{s=1}^S \lvert \tilde{x}_s(t;\mathbf{k}^\text{MLE},\mathbf{x}(0)) - \tilde{x}_s(t;\mathbf{k}^\text{true},\mathbf{x}(0)) \rvert \, dt.
\end{equation}
We compute the integral in \eqref{eq:err} numerically over a grid of 1000 evenly spaced time points. Trajectory prediction errors are defined similarly, using the novel initial state $\mathbf{x}(0)=(1,0,0)$.

\begin{figure}[t]
\centering
\includegraphics[width = \textwidth]{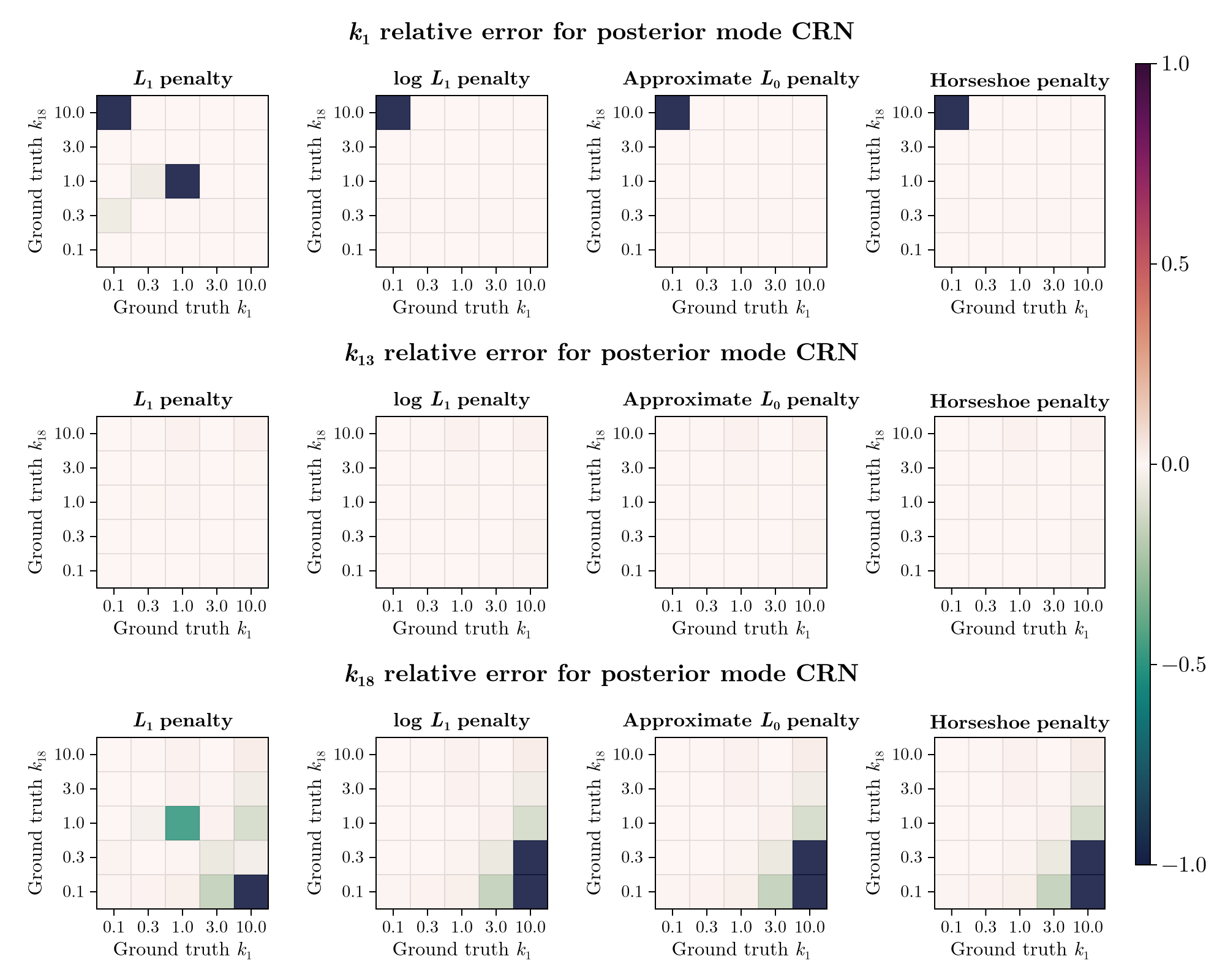}
\caption{Relative errors of rate constant MLEs of the posterior mode CRNs obtained with each penalty function for the simulation study.}\label{fig:toy_k_err}
\end{figure}

In Figure~\ref{fig:toy_k_err}, we show the relative errors of $\mathbf{k}^\text{MLE}$ of the posterior mode CRNs obtained with each penalty function. The relative error of the rate constant of a ground-truth reaction $r$ is
\begin{equation*}
    \frac{k^\text{MLE}_r - k^\text{true}_r}{k^\text{true}_r}.
\end{equation*}
A relative error of $-1$ for reaction $r$ occurs when $k^\text{MLE}_r=0$, in other words, the posterior mode CRN does not include reaction $r$.

Figure~\ref{fig:toy_top_crns} is the simulation study counterpart to Figures~\ref{fig:pinene}B and \ref{fig:pyridine}B of the main text.

\begin{figure}[p]
\centering
\includesvg[width=\textwidth, inkscapearea=page, inkscapelatex=false]{figures/toy_top_crns.svg}
\caption{Unnormalised log posterior of the top 20 CRNs pooled from the CRN ensembles $\CRNset(\estset)$ obtained with each penalty function for the simulation study. Below each line plot, the shaded cells indicate whether a CRN is present in the CRN ensemble $\CRNset(\estset)$ obtained with some penalty function. Starred cells correspond to CRNs that are present in $\CRNset(\estset)$ (after recombination) but not in $\CRNset_\text{base}(\estset)$ (before recombination).}\label{fig:toy_top_crns}
\end{figure}

\clearpage

\subsection{Case study 1: \texorpdfstring{\textalpha}{α}-pinene isomerisation}

\begin{figure}[h!]
\centering
\includegraphics[width = 0.6\textwidth]{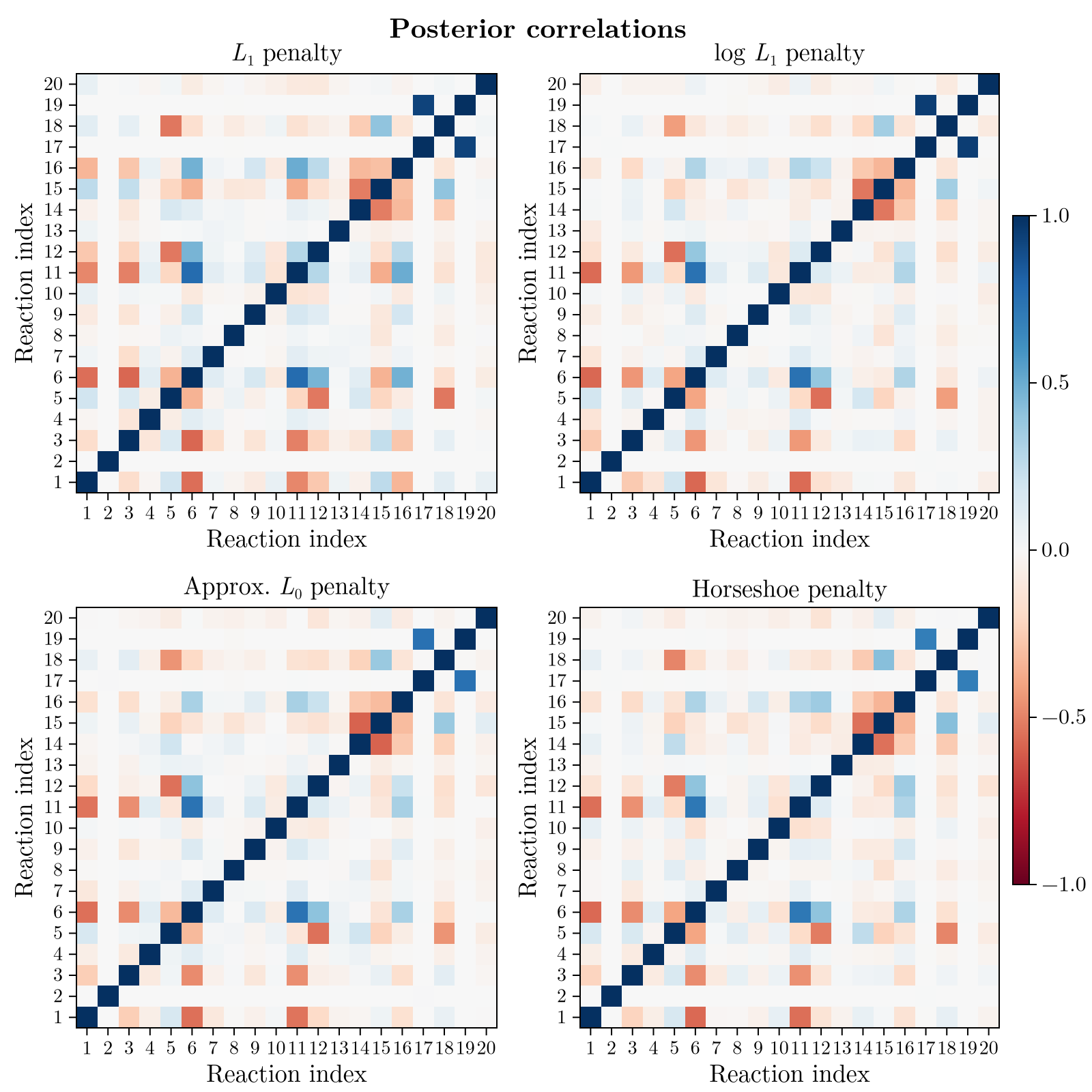}
\caption{Posterior correlations of reaction inclusion for the \textalpha-pinene isomerisation case study.}\label{fig:pinene_corr}
\end{figure}

From Figure~\ref{fig:pinene_corr}, we see that the posterior correlations obtained with each penalty function are qualitatively similar to each other.

We visualise the hierarchical representation of the 95\% HPD set $\CRNset^\text{HPD}$ obtained with the $\log L_1$ penalty in Figure~\ref{fig:pinene_hpd_tree}. All CRNs in $\CRNset^\text{HPD}$ include the reactions $X_1 \xrightarrow{k_{2}} X_3$, $2X_1 \xrightarrow{k_{17}} X_5$, $2X_3 \xrightarrow{k_{19}} X_5$, which all appear in the CRN proposed in \cite{stewart_bayesian_1981}. Interestingly, $2X_1 \xrightarrow{k_{17}} X_5$ was not proposed in the original work by \textcite{fuguitt_liquid_1945}, yet its inclusion in all CRNs of $\CRNset^\text{HPD}$ suggests that this reaction is essential in explaining the observed data. Recall that each tree node $v$ is associated with a set of CRNs $\CRNset(v)$, characterised by the inclusion of reactions $\CRN^\text{inc}(v)$ and the exclusion of reaction $\CRN^\text{exc}(v)$.  For the leaf node $v$ with the highest posterior probability, we have $\CRN^\text{inc}(v)=\{2,17,19,3,1,5\}$. The CRN set $\CRNset(v)$ consists of the CRN $\CRN^\text{inc}(v)$ and 51 other CRNs that include $\CRN^\text{inc}(v)$ as a subnetwork, indicated by the lack of reactions in the node $v$ label (see Section~S2.4). 

Let $v_{3,1}, v_{3}, v_{1}$ be the nodes characterised by 
\begin{alignat*}{2}
    \CRN^\text{inc}(v_{3,1}) &= \{2,17,19,3,1\},\qquad & \CRN^\text{exc}(v_{3,1})&=\varnothing, \\
    \CRN^\text{inc}(v_{3}) &= \{2,17,19,3\},\qquad & \CRN^\text{exc}(v_{3})&=\{1\}, \\
    \CRN^\text{inc}(v_{1}) &= \{2,17,19,1\},\qquad & \CRN^\text{exc}(v_{1})&=\{3\}. 
\end{alignat*}
These three nodes (coloured blue) are ordered from left to right along the tree. Every CRN in $\CRNset^\text{HPD}$ belongs to one of the subtrees rooted at $v_{3,1}, v_{3}, v_{1}$. This implies that the CRNs in $\CRNset^\text{HPD}$ must include at least one of the reactions $X_1 \xrightarrow{k_{1}} X_2$ or $X_1 \xrightarrow{k_{3}} X_4$. Moreover, all CRNs belonging to the subtrees rooted at $v_{3}$ and $v_{1}$ include reactions $X_2 \xrightarrow{k_{6}} X_4$ and $X_4 \xrightarrow{k_{11}} X_2$. The same is not true for the subtree rooted at $v_{3,1}$, i.e. CRNs in this subtree do not necessarily include reactions $X_2 \xrightarrow{k_{6}} X_4$ and $X_4 \xrightarrow{k_{11}} X_2$. In summary, every CRN in $\CRNset^\text{HPD}$ includes one of the following three reaction schemes:
\begin{equation*}
    X_2 \leftarrow X_1 \rightarrow X_4, \qquad 
    X_1 \rightarrow X_2 \rightleftharpoons X_4, \qquad
    X_1 \rightarrow X_4 \rightleftharpoons X_2.
\end{equation*}

Upon inspecting each of the three subtrees, we notice that every CRN in $\CRNset^\text{HPD}$ must include at least one of the three reactions $X_2 \xrightarrow{k_{5}} X_3$, $X_4 \xrightarrow{k_{12}} X_3$, $X_5 \xrightarrow{k_{15}} 2X_3$. In other words, the reaction $X_1 \xrightarrow{k_{2}} X_3$, which all CRNs in $\CRNset^\text{HPD}$ include, is insufficient to explain the observed production of $X_3$ solely by itself.

For CRNs belonging to the subtrees rooted at $v_{3}$ and $v_{1}$, we note that when a CRN excludes both $X_2 \xrightarrow{k_{5}} X_3$ and $X_4 \xrightarrow{k_{12}} X_3$, there is a conversion of $X_5$ to $X_3$. In this case, the CRN also includes at least one of $2X_2 \xrightarrow{k_{18}} X_5$ or $2X_4 \xrightarrow{k_{20}} X_5$ (bottom row of leaf nodes). In other words, there is an additional reaction that produces $X_5$ apart from $2X_1 \xrightarrow{k_{17}} X_5$ or $2X_3 \xrightarrow{k_{19}} X_5$.

For CRNs belonging to the subtree rooted at $v_{3,1}$, they all include at least one of the three reaction pathways $X_2 \xrightarrow{k_{5}} X_3$, $2X_2 \xrightarrow{k_{18}} X_5 \xrightarrow{k_{15}} 2X_3$, $X_2 \xrightarrow{k_{6}} X_4 \xrightarrow{k_{12}} X_3$. These are pathways that convert $X_2$ to $X_3$, possibly with $X_4$ or $X_5$ as intermediate species.

In conclusion, we have extracted higher-order information about network uncertainty from a tree visualisation that cannot simply be deduced from posterior correlations. However, the current approach is quite reliant on manual inspection of the tree. Automating the hierarchical analysis of network uncertainty is a future direction of research.

\begin{figure}[p]
\centering
\includesvg[width=\linewidth, inkscapearea=page, inkscapelatex=false]{figures/pinene_hpd_tree_logL1.svg}
\caption{Hierarchical representation of the 95\% HPD set obtained with the $\log L_1$ penalty for the \textalpha-pinene isomerisation example. The root node represents the 95\% HPD set, which is sequentially split according to reaction inclusion as indicated by the edges. Each node represents a subset of the 95\% HPD set, labelled with its posterior probability and number of CRNs in brackets. Node marker size scales with the number of CRNs. Node colour is referenced in Section S4.2. See Section~S2.4 for how splits are determined and details about the leaf nodes. The inset lists the candidate reactions.}\label{fig:pinene_hpd_tree}
\end{figure}

\clearpage

\subsection{Case study 2: pyridine denitrogenation}

\begin{figure}[h!]
\centering
\includegraphics[width = 0.9\textwidth]{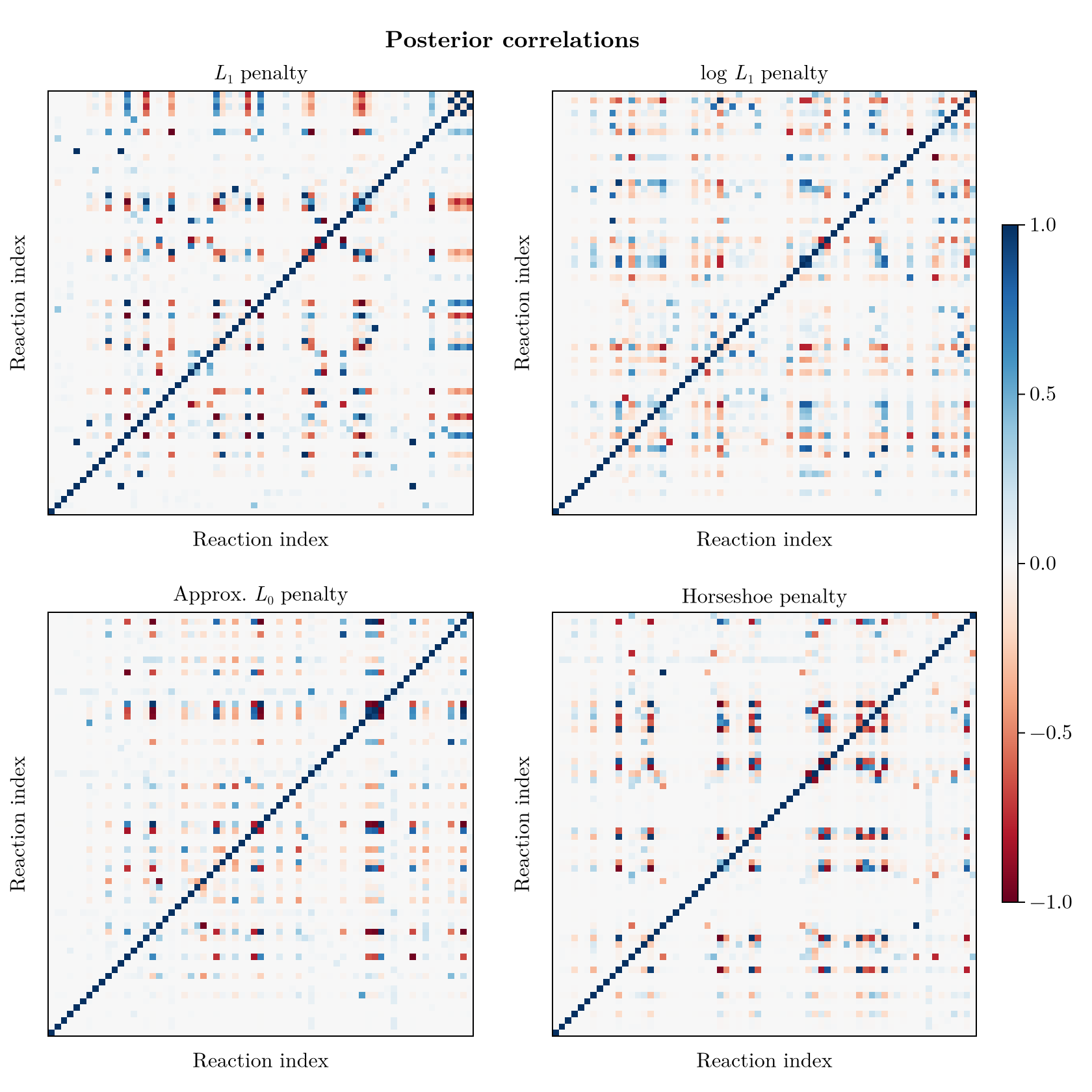}
\caption{Posterior correlations of reaction inclusion for the pyridine denitrogenation case study.}\label{fig:pyridine_corr}
\end{figure}

From Figure~\ref{fig:pyridine_corr}, we see that the posterior correlations obtained with each penalty function are quite different. This suggests that the CRNs with the highest posterior probabilities have not been robustly found by any single penalty function.

We compute the MLE for the gold-standard CRN $\CRN^\text{gold}$ (11 reactions) as described in Section~\ref{sec:pyridine_lib}. Let $k^\text{gold}_r$ denote the rate constant estimate for reaction $r$. For each gold-standard reaction $r$, we report its dimensionless rate constant estimate and reaction contribution in Table~\ref{tbl:pyridine_gold}, which indicate how important reaction $r$ is to the observed dynamics.

We visualise the hierarchical representation of the 95\% HPD set $\CRNset^\text{HPD}$ obtained with the $\log L_1$ penalty in Figure~\ref{fig:pyridine_hpd_tree}. The aim of this section is to explain why the gold-standard CRN is only partially recovered. The only gold-standard reaction not featured in Figure~\ref{fig:pyridine_hpd_tree} is reaction $10$. There are five gold-standard reactions present in all CRNs in $\CRNset^\text{HPD}$, namely reactions $1$, $5$, $9$, $12$, $46$. These reactions are also the gold-standard reactions with the highest contributions (Table~\ref{tbl:pyridine_gold}), indicating that there is more evidence in the observed data for these gold-standard reactions. The gold-standard reaction with the sixth-highest reaction contribution is reaction $61$, which has a high posterior probability of 0.87 (not shown). 

The remaining five gold-standard reactions yet to be mentioned (namely $4$, $10$, $17$, $32$, $64$) have relatively low reaction contributions, and do not have a significant presence in Figure~\ref{fig:pyridine_hpd_tree}. The dynamics induced by these reactions are either insignificant to the observed dynamics, or can be explained by other reactions in Figure~\ref{fig:pyridine_hpd_tree}. For instance, reaction $4$, $X_2\rightarrow X_1$ has a contribution 1-2 orders of magnitude smaller than the observed range of $X_1$ and $X_2$, and thus has little influence on the observed dynamics. Consider the gold-standard reaction $64$, $X_5+X_6\rightarrow 2X_3$, as another example. Note that reaction $58$, $X_4+X_6\rightarrow 2X_3$, which is included by all CRNs in $\CRNset^\text{HPD}$, replaces $X_5$ in reaction $64$ with $X_4$, and that the observed trajectories of $X_4$ and $X_5$ share very similar contours. Since the contribution of reaction $64$ is relatively low (Table~\ref{tbl:pyridine_gold}), there is likely insufficient signal in the data to distinguish between $X_4+X_6\rightarrow 2X_3$ and $X_5+X_6\rightarrow 2X_3$. 



In conclusion, it is likely that the difference between the 95\% HPD set and the gold-standard CRN can be explained by a lack of information present in the observed dynamics. However, we stress that the absence of the gold-standard CRN from the 95\% HPD set indicates that network uncertainty is underestimated for this case study.

\begin{table}[t]
\centering\small
\begin{tabular}{*{4}{c}}
\toprule
$r$ & Reaction $r$ & $k^\text{gold}_r/\kappa_r$ & $g(r, \mathbf{k}^\text{gold})$ \\
\midrule
1 & $X_1 \rightarrow X_2$ & 1.8068 & 1.0140\\
4 & $X_2 \rightarrow X_1$ & 0.0055 & 0.0146\\
5 & $X_2 \rightarrow X_3$ & 0.3469 & 0.8745\\
9 & $X_3 \rightarrow X_6$ & 0.1909 & 0.3340\\
10 & $X_4 \rightarrow X_5$ & 0.0446 & 0.0336\\
12 & $X_4 \rightarrow X_2$ & 0.1511 & 0.4559\\
17 & $X_5 \rightarrow X_3$ & 0.4529 & 0.0546\\
32 & $2X_3 \rightarrow X_5 + X_6$ & 0.0274 & 0.0353\\
46 & $X_2 + X_3 \rightarrow X_4 + X_6$ & 0.3178 & 0.7492\\
61 & $X_4 + X_6 \rightarrow X_2 + X_3$ & 0.0734 & 0.2165\\
64 & $X_5 + X_6 \rightarrow 2X_3$ & 0.0122 & 0.0134\\
\bottomrule
\end{tabular}
\tblspace
\caption{Dimensionless rate constant estimates and reaction contributions~\eqref{eq:reaction_flux} for each reaction of the gold-standard CRN for the pyridine denitrogenation example. Note that $r$ represents the reaction indices introduced in Section~\ref{sec:pyridine_lib}.}\label{tbl:pyridine_gold}
\end{table}

\clearpage

\begin{figure}[p]
\centering
\includesvg[width=\linewidth, inkscapearea=page, inkscapelatex=false]{figures/pyridine_hpd_tree_logL1.svg}
\caption{Hierarchical representation of the 95\% HPD set obtained with the $\log L_1$ penalty for the pyridine denitrogenation example. The root node represents the 95\% HPD set, which is sequentially split according to reaction inclusion as indicated by the edges. Each node represents a subset of the 95\% HPD set, labelled with its posterior probability and number of CRNs in brackets. Node marker size scales with the number of CRNs. See Section~S2.4 for how splits are determined and details about the leaf nodes. The inset lists the candidate reactions. Bolded reaction numbers refer to reactions present in the gold-standard CRN.}\label{fig:pyridine_hpd_tree}
\end{figure}

\clearpage

\clearpage
\AtNextBibliography{\small}
\emergencystretch 2em
\printbibliography[heading=bibintoc]

\end{refsection}

\end{document}